\documentclass[12pt,preprint]{aastex}

\shorttitle{Circumstellar interaction in SNe~IIP}
\shortauthors{Chugai N.N. et al.}

\begin{document}

\title{OPTICAL SIGNATURES OF CIRCUMSTELLAR INTERACTION IN TYPE IIP SUPERNOVAE}

\author{
Nikolai N. Chugai\altaffilmark{1,2},
Roger A. Chevalier\altaffilmark{2},
Victor P. Utrobin\altaffilmark{3}
}
\altaffiltext{1}{Institute of Astronomy, RAS, Pyatnitskaya 48, 109017 Moscow, Russia;
nchugai@inasan.ru}
\altaffiltext{2}{Department of Astronomy, University of Virginia, P.O. Box 400325, 
Charlottesville, VA 22904; rac5x@virginia.edu}
\altaffiltext{3}{Institute of Theoretical and Experimental Physics, 117218 Moscow,
Russia; utrobin@itep.ru}

\begin{abstract}

We propose new diagnostics for circumstellar interaction
in Type IIP supernovae (SNe~IIP) 
by the detection of high velocity (HV) absorption features
in H$\alpha$ and He\,I 10830 \AA\ lines during the photospheric stage.
To demonstrate the method, we compute the ionization and excitation
of H and He in supernova ejecta taking into account
time-dependent effects and X-ray irradiation.
We find that the interaction with a typical red supergiant wind should
result in the enhanced excitation of the 
outer layers of unshocked ejecta and the emergence of  corresponding 
HV absorption, i.e. a depression in the blue
absorption wing of H$\alpha$ and a
pronounced absorption of He\,I 10830 \AA\ at a radial velocity
of about $-10^4$ km s$^{-1}$. We identify HV absorption 
in H$\alpha$ and He\,I 10830 \AA\ lines of SN~1999em and
in H$\alpha$ of SN~2004dj as being due to this effect. The derived
mass loss rate is close to $10^{-6}~M_{\odot}$ yr$^{-1}$ for both supernovae,
assuming a wind velocity 10 km s$^{-1}$.
We argue that, in addition to the HV absorption formed in the unshocked ejecta,
spectra of SN~2004dj and SN~1999em show a HV notch feature that is
formed in the cool dense shell (CDS) modified by the Rayleigh-Taylor instability.
The CDS results from both shock breakout and 
radiative cooling of gas that has passed through the reverse shock wave.
The notch becomes dominant in the HV absorption during the late photospheric 
phase, $\ga60$ d. The wind density deduced from the velocity of the CDS 
is consistent with the wind density found from the HV absorption 
produced by unshocked ejecta.

\end{abstract}


\section{INTRODUCTION} \label{sec-intro}

The link between a certain Type IIP supernova (SN~IIP) and its main-sequence
progenitor is poorly known despite the widely accepted view that these
supernovae arise from the initial mass range of $10-25~M_{\odot}$ \citep[e.g.,][]{Heg03}.
The primary reasons for the uncertainty in this field
 are the small number of
presupernova (pre-SN) mass determinations and the uncertain amount of matter
lost through winds, presumably dominated by the
red supergiant (RSG) wind. While the former is becoming
well-studied, the latter uncertainty is related to
our poor knowledge of the complicated physics of mass loss and the unsatisfactory
situation with the empirical measurements of
mass loss from  RSGs. Even for the most studied close massive
RSG,  Betelgeuse, the range of observational estimates is large:
$2\times10^{-7}-1.5\times10^{-5}~M_{\odot}$ yr$^{-1}$ \citep{JK90,Loon05}.

The situation for SNe~IIP has some promise because 
the dominant mass loss at
the RSG stage may be observed by the detection of circumstellar (CS)
interaction at X-ray and radio wavelengths.  
The Type IIP SN~1999em
was detected in X-rays with {\em Chandra} \citep{Poo02}, leading to the 
mass loss estimate $\dot{M}\approx(1-2)\times10^{-6}~M_{\odot}$ yr$^{-1}$
(assuming a wind velocity of 10 km s$^{-1}$).
A recent study of available X-ray and radio data for SNe~IIP
produced a range of mass loss rates of pre-SN~IIP
 $\dot{M}\sim(1-10)\times10^{-6}~M_{\odot}$ yr$^{-1}$
\citep{CFN06}. The application of these rates to the full RSG stage ($\sim10^6$ yr)
suggests the loss of $1-10~M_{\odot}$ for pre-SNe~IIP.
The wide range of estimates emphasizes the need for the individual 
determination of
the wind density for each particular SN~IIP under consideration. 
Unfortunately, this is not always possible since X-ray and
radio observations of SNe~IIP are often not available.

Here, we propose two new diagnostics
for the wind density in SNe~IIP that could help. Both rely on
spectroscopic observations of H$\alpha$ and He\,I 10830 \AA\ at the
photospheric epoch. The first one is based on the
fact that the interaction of SN ejecta with the wind results in
the emission of X-rays from both forward and reverse shocks.
The X-rays cause ionization and excitation of
SN ejecta that may be revealed, e.g., through specific emission lines \citep{CF94}.
Unfortunately, in SNe~IIP the wind density is low and
emission lines caused by CS interaction are extremely weak and 
cannot be detected.
We find, however, that the excitation of H and He
produced by X-rays in SN~IIP ejecta turns out to be
sufficient to be detected as high velocity (HV) absorption features in H$\alpha$
and He\,I 10830 \AA\ lines against the bright SN~IIP photosphere.
This is the core of our proposed diagnostic for the wind in SNe~IIP.
The second proposed probe for CS interaction exploits the possibility
that a cool dense shell (CDS) might form at the SN/CS interface
because of radiative cooling.
The CDS excited by X-rays could become visible as narrow HV absorption
in H$\alpha$. The velocity of this absorption would
provide a direct measure of the expansion velocity of
the SN/CS interface, a valuable dynamical characteristic of the
CS interaction.

The identification of the expected HV lines in observed spectra
is complicated by the presence of weak metal lines \citep{DH05}.
However, we argue that HV lines of H$\alpha$ and
H$\beta$ have been observed in spectra of SNe~IIP, 
as previously discussed by \cite{Leo02}
for SN~1999em.

The paper is organized as follows. We start with a description
of the interaction model and CS interaction effects
in the H$\alpha$ and He\,I 10830 \AA\ absorption lines formed in the
unshocked ejecta during the photospheric epoch (\S~\ref{sec-mod}).
We then compare our CS interaction models
with the available spectra of SNe~IIP in the H$\alpha$ and
He\,I 10830 \AA\ lines and estimate
the wind density for particular SNe~IIP  (\S~\ref{sec-inter}).
In \S~\ref{sec-narrow} we address the issue of
the H$\alpha$ absorption in the cool dense shell
at the SN/CS interface of SNe~IIP. 
We discuss 
 implications of our models in the last section.


\section{MODEL}\label{sec-mod}

The model for ejecta-wind interaction effects in
H$\alpha$ and He~I 10830 \AA\ lines formed in the SN ejecta
consists of three major parts:
(i) an interaction model that provides the dynamical evolution
 of the SN/wind interface and
the X-ray emission from the reverse and forward shocks;
(ii) a model for the ionization and
excitation of H and He in the unshocked SN ejecta irradiated
by X-rays; and (iii) the calculation of line profiles.
We perform (iii) using either a standard Sobolev approximation or the
direct integration of the radiation transfer equation, depending on the
validity of the Sobolev approximation.

\subsection{The interaction model}\label{sec-interh}

The interaction of SN ejecta with CS wind leads to a canonical 
double-shock structure \citep{Che82a,Nad85}
with the forward shock propagating in the CS 
gas and the reverse shock in the SN ejecta.
We treat the CS interaction of ejecta in 
the thin shell approximation \citep{Che82b} in which the
double-shock layer is reduced to an infinitely thin shell. 
We assume that the freely
expanding ($v=r/t$) SN envelope has a sharp boundary at the velocity
$v_{\rm c}$ and begins to interact with a wind starting 
some moment ($t\sim1$ day)
which corresponds roughly to the shock breakout phase.
Free expansion is expected to take several doubling times of the
initial radius, or several days, to be set up, but the observations
we are modeling are at later times.
The maximum velocity $v_{\rm c}$ is set by the escape of radiation from the
shock wave at shock breakout.

For the density distribution $\rho(v)$ in the SN~IIP envelope we use
an analytical expression,
\begin{equation}
\rho=\frac{\rho_0}{1+(v/v_0)^k}\,,
\label{eq-den}
\end{equation}
that closely approximates the combination of an inner plateau and outer
power law tail found in hydrodynamic models
\citep[e.g.,][]{U07}.
The parameters $\rho_0$ and $v_0$ are determined by the kinetic energy $E$,
ejecta mass $M$, and $k$:
\begin{equation}
M=4\pi \rho_0(v_0t)^3C_{\rm m}\,,   \qquad
E=\frac{1}{2}\frac{C_e}{C_m}Mv_0^2 \,,
\label{eq-v0}
\end{equation}
where
\begin{equation}
C_{\rm m}=\frac{\pi}{k\sin(3\pi/k)}\,, \qquad C_{\rm e}=\frac{\pi}{k\sin(5\pi/k)}\,.
\label{eq-cmce}
\end{equation}
The power law index $k$ lies in the range $k\sim 8-10$ and is henceforth 
set to be $k=9$.
We adopt the boundary velocity for a typical SN~IIP of
$v_{\rm c}=1.5\times10^4$ km s$^{-1}$, which is consistent with
models of shock breakout \citep{MM99,CFN06}.
Also, the blue edge of the H$\beta$
emission in SN~1999gi on day 1 was observed
to be at 15,000 km s$^{-1}$ \citep{LFW02}.

Hydrodynamical modeling predicts that at the shock breakout phase
a thin dense shell forms at the outer boundary
\citep{GIN71,FA77,Che81,U07}. The physics
of the `boundary shell' formation is in the transition from the
adiabatic to the radiative
regime of  shock wave propagation in the outermost layers
of the exploding star.
Simple considerations of radiative diffusion \citep{Che81} give
an estimate for the shell mass:
\begin{equation}
M_s\approx 2\times 10^{-4}
\left(R\over 500~R_{\odot}\right)^2
\left(v_b\over 10,000~{\rm km~s^{-1}}\right)^{-1}
\left(\kappa \over 0.4~{\rm cm^2~g^{-1}}\right)^{-1}  ~M_{\odot},
\end{equation}
where $R$ is the progenitor star radius and $v_b$ is the velocity
at shock breakout, before free expansion has been established.
During the acceleration phase the boundary shell is subject to
the Rayleigh-Taylor (RT) instability \citep{FA77} and therefore
could be corrugated or even fragmented. However, it is not
clear whether the RT instability results in  full fragmentation; 
 2D or 3D hydrodynamic modeling of the phenomenon is lacking.
We will treat the boundary shell in our model as an intact thin 
spherical shell which is already in place at the initial moment.
The mass of the boundary shell is presumably equal to the mass
of the outer envelope ($v>v_{\rm c}$) with an extrapolated power
law $\rho\propto v^{-k}$. This provides a good estimate of the
shell mass for SN~1999em, $\sim3\times10^{-4}~M_{\odot}$,
in agreement with hydrodynamical modeling \citep{U07}.
The boundary shell is considered below as a seed for the CDS that may,
or may not, grow further due to radiative cooling at the reverse shock.

To calculate the X-ray emission from the reverse and forward shocks
 we assume that the postshock
layers are uniform and their densities are
 4 times larger than the corresponding preshock density.
We allow the electron temperature to be partially equilibrated,
assuming the following interpolation:
\begin{equation}
T_e=\mbox{max}\left(0.1T_{\rm eq},\,\frac{t}{t+t_{\rm eq}}T_{\rm eq}\right)\,,
\end{equation}
where $t_{\rm eq}$ is the electron-ion temperature equilibration time
\citep{Spitz62} and $T_{\rm eq}$ is the equilibrium shock temperature
with $T_e=T_i$. The lower limit $T_e/T_{\rm eq}=0.1$ is indicated by
observational data on supernova remnants with shock velocities of
$\sim 1000$ km s$^{-1}$ \citep{Rak05}. The
reverse shock is equilibrated during the photospheric phase for a 
mass loss rate of $\sim 10^{-6}~M_{\odot}$ yr$^{-1}$ with a
 wind velocity of $10$ km s$^{-1}$,
while the forward shock is not. Since the reverse shock is equilibrated and
dominates the X-ray luminosity, the uncertainty in $T_e$ does 
not affect our results. The shock X-ray luminosity is expressed in terms of the
kinetic luminosity as $L_{\rm X}=\eta_{\rm X}L_{\rm kin}$.
The efficiency parameter $\eta_{\rm X}$ is smoothly interpolated between
extreme cases, $\eta_{\rm X}=t/(t+t_{\rm c})$, where
 $t_{\rm c}$ is the cooling time and
is calculated for each shock using the cooling
function of \cite{SD93}.
The wind is assumed to be steady with the density 
 $\rho \propto r^{-2}$.
Since all the interaction effects for low wind velocities ($u$)
are determined by the wind density $\rho\propto \dot{M}/u$, 
we use hereafter the dimensionless wind 
density parameter $w=\dot{M}_{-6}/u_{10}$,
where $\dot{M}_{-6}$ is the mass loss rate in units of $10^{-6}~M_{\odot}$ yr$^{-1}$ and $u_{10}$ is the wind velocity in units of $10$ km s$^{-1}$.

\begin{figure}
\plotone{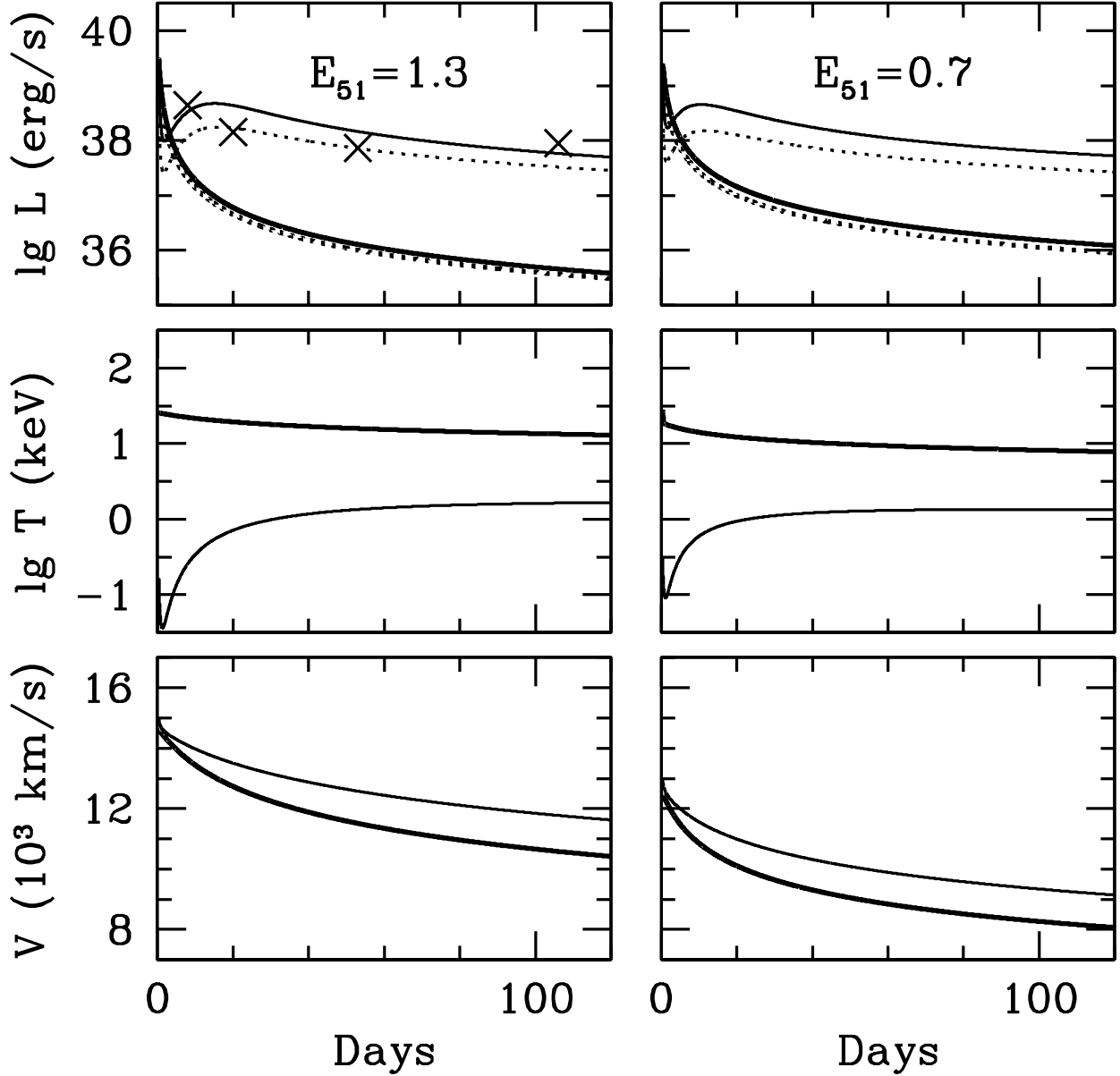}
\caption{Major properties of the interaction model for two
values of the kinetic energy of the SN ejecta, $1.3\times10^{51}$ erg ({\em left})
and $0.7\times10^{51}$ erg. The {\em top} panel shows the X-ray
luminosity of the forward ({\em thick} lines) and reverse ({\em thin})
shocks for unabsorbed ({\em solid}) and absorbed ({\em dotted})
cases. In the left upper panel, {\em Chandra} X-ray luminosity 
measurements of SN~1999em
\citep{Poo02} are plotted by {\em crosses}.
The {\em middle} panel displays the electron temperature of
the forward ({\em thick} lines) and reverse ({\em thin})
shocks.  The {\em lower} panel shows the
velocity of the thin shell ({\em thick}  line) and boundary velocity of
the unshocked SN ejecta ({\em thin}  line). In both cases the X-ray
luminosity is dominated by the reverse shock.
 }
  \label{f-dyn}
  \end{figure}

The interaction model for a SN~IIP with the wind density $w=1$
is illustrated in Fig. \ref{f-dyn}. The two cases  shown 
have the same ejecta mass $M=19~M_{\odot}$ but 
different kinetic energy:
$E=1.3\times10^{51}$ erg and $E=0.65\times10^{51}$ erg.
The high energy case corresponds presumably to SN~1999em 
\citep{U07}, while the low energy case 
illustrates SN~2004dj, as we will see below.
Accordingly, we adopt $v_{\rm c}=15,000$ km s$^{-1}$ in the 
high energy case and $v_{\rm c}=13,000$ km s$^{-1}$ in the low energy case.
The model with parameters for SN~1999em reproduces the observed 
(absorbed) {\em Chandra} X-ray
luminosity of SN~1999em in the $0.5-8$ keV band 
\citep{Poo02} for a distance $D=11.7$ Mpc \citep{Leo03}
quite satisfactorily (Fig. \ref{f-dyn}), 
 which indicates that $w=1$ is a reasonable choice for the wind density.
The estimated  
Galactic absorption, $N_{\rm H}=6\times10^{20}$ cm$^{-2}$ \citep{Poo02}, is small 
and does not significantly affect  the observed X-ray luminosity.
The X-ray luminosity from the reverse shock dominates in both cases.
 The typical electron temperature for the reverse shock
at $t>20$ d is $\sim1$ keV, while for the forward shock it is $\sim10$ keV.
The early behavior of the velocities
differs from the self-similar evolution \citep{Che82b} 
because of the ejecta density cut-off in our model and the
presence of the boundary shell at the initial moment.
The mass of the boundary shell is $3\times10^{-4}~M_{\odot}$
and $1.1\times10^{-4}~M_{\odot}$ in the high and low energy models,
respectively. The total
mass of the CDS attained at the time of the termination of the
radiative regime in the reverse shock ($t_{\rm rad}$)
is  $6\times10^{-4}~M_{\odot}$ and $3.2\times10^{-4}~M_{\odot}$
with $t_{\rm rad}=21$ d  and $t_{\rm rad}=18$ d
 for the high and low energy models, respectively.
Remarkably, the reverse shock is adiabatic
in the self-similar models for the same parameters
 through this epoch. This fact 
is related to the different dynamics of the self-similar 
model at an early stage, which
results in a lower preshock density at the reverse shock.

The spectrum of X-rays is described below using a standard 
approximation $F_E\propto E^{-q}\exp(-E/T_e)$. We adopt $q=0.4$
for the forward shock, a reasonable approximation for temperature
$\sim10$ keV \citep{Cox00}, while for the
low temperature reverse shock we take $q=1$. The latter choice
qualitatively describes the calculated energy distribution of X-rays
with a significant contribution of emission lines for
the electron temperature $T_e=1$ keV \citep{NFK06}.

\subsection{Model of H and He excitation}\label{sec-profile}

The hydrogen and helium ionization in a SN~IIP atmosphere at the photospheric epoch
is time-dependent \citep{UC02,UC03,UC05}. To take this effect into account,
we use a hydrogen atom represented by four levels and continuum.
All the essential radiative and collisional processes for hydrogen 
are taken into account,
including non-thermal ionization and excitation induced by X-ray absorption.
The excitation of the helium $2^3$S level is determined by a simple model with
two excited levels (2$^1$S and 2$^3$S).
We consider only major processes for the population of the 2$^3$S level:
 nonthermal excitation and ionization with subsequent recombination
into triplet states, depopulation by the collisional deexcitation
transition $2^3$S--$2^1$S, Penning ionization \citep{Chu91} and
continuum absorption in the 10830 \AA\ line (2$^3$S--2$^3$P transition)
with the subsequent spontaneous transition 2$^3$P--1$^1$S terminated by
photon escape. The energy balance
includes the energy deposition due to the absorption of X-rays
and the hydrogen photoionization from excited levels by photospheric radiation.
The cooling includes adiabatic loss ($pdV$ work), free-free,
free-bound, bound-bound emission
of hydrogen, and cooling in the [O\,I] 6300, 6364 \AA\ and  Mg\,II 2800 \AA\ lines. The
Fe II line cooling is approximately included 
by multiplying the Mg\,II 2800 \AA\ cooling rate by a factor of two.
Given the possibility of a high temperature regime ($T_e>20,000$~K)
in the outer layers, the cooling at this range of
temperatures is treated using the cooling function of \cite{SD93}.

The distribution of the ionization and excitation of H and He\,I in the
atmosphere at one time is calculated by tracking the evolution
of a Lagrangian zone after its emergence from the photosphere.
The initial ionization and excitation at the photosphere is set
assuming Saha-Boltzmann equations for the effective temperature.
Subsequent ionization and excitation are computed by
solving the system of time-dependent
kinetic equations together with the time-dependent energy balance.
The average intensity of the photospheric radiation is assumed to be
$J_{\nu}=WI_{\nu}$, where $W$ is the dilution factor and $I_{\nu}$ is
the photospheric brightness. The
luminosity evolution at the photospheric epoch ($t<100$ d) is
described by the expression
\begin{equation}
L=L_0[1+\exp(-t/t_1)]\,.
\end{equation}
With $L_0=10^{42}$ erg s$^{-1}$ and $t_1=10$ d, this
corresponds to the bolometric luminosity of SN~1999em
\citep{Elm03}, assuming a distance of 11.7 Mpc. The photospheric radius 
is determined by the velocity at the photosphere and the age 
($r_{\rm p}=v_{\rm p}t$), where
$v_{\rm p}$ is deduced from observational data of \cite{Leo02}.
The spectrum  of the photospheric continuum is taken to be blackbody with
  $T=T_{\rm eff}$ in the wavelength region longward of the Balmer jump,
  while the ultraviolet (UV) radiation
 shortward of the Balmer continuum edge is suppressed by a factor
 $f_{\rm UV}\leq1$ which was assumed to monotonically
drop after the explosion from unity to 0.03 on a timescale
of 10 d. 
This evolution was deduced from the behavior observed in SN~1987A
\citep{Pun95}. We found that the results are only weakly sensitive 
to the specific form of the variation of $f_{\rm UV}$. The energy 
deposition rate in the
atmosphere due to the absorption of X-rays is calculated with
an absorption coefficient $k_{\rm X}=100E^{-8/3}$ cm$^2$ g$^{-1}$ 
(where $E$ is in keV).
The deposited energy is shared between heating,
ionization, and excitation according to the recipes of \cite{KF92}
and \cite{Xu92}.

 The solution of the kinetic equations for each Lagrangian zone
provides populations
 of the hydrogen levels $n_2$ and $n_3$, and the $2^3$S level of
He\,I in the atmosphere. These
 values are then used to calculate line profiles of H$\alpha$
and He\,I 10830 \AA. We compared our model in the case of zero wind
density with the detailed model of \cite{UC05} and found that
the present model results in slightly weaker H$\alpha$ absorption;
at the relative intensity of 0.9, the radial velocity in the simple model is lower
by $\sim500$ km s$^{-1}$. This drawback of the simple model
does not noticeably affect the amplitude of the calculated HV absorption formed
in the external layers, where the ionization unaffected
by X-ray absorption is small.


\section{HIGH VELOCITY FEATURES IN H AND He LINES}\label{sec-inter}


\subsection{Model properties}\label{sec-property}

\begin{figure}
\plotone{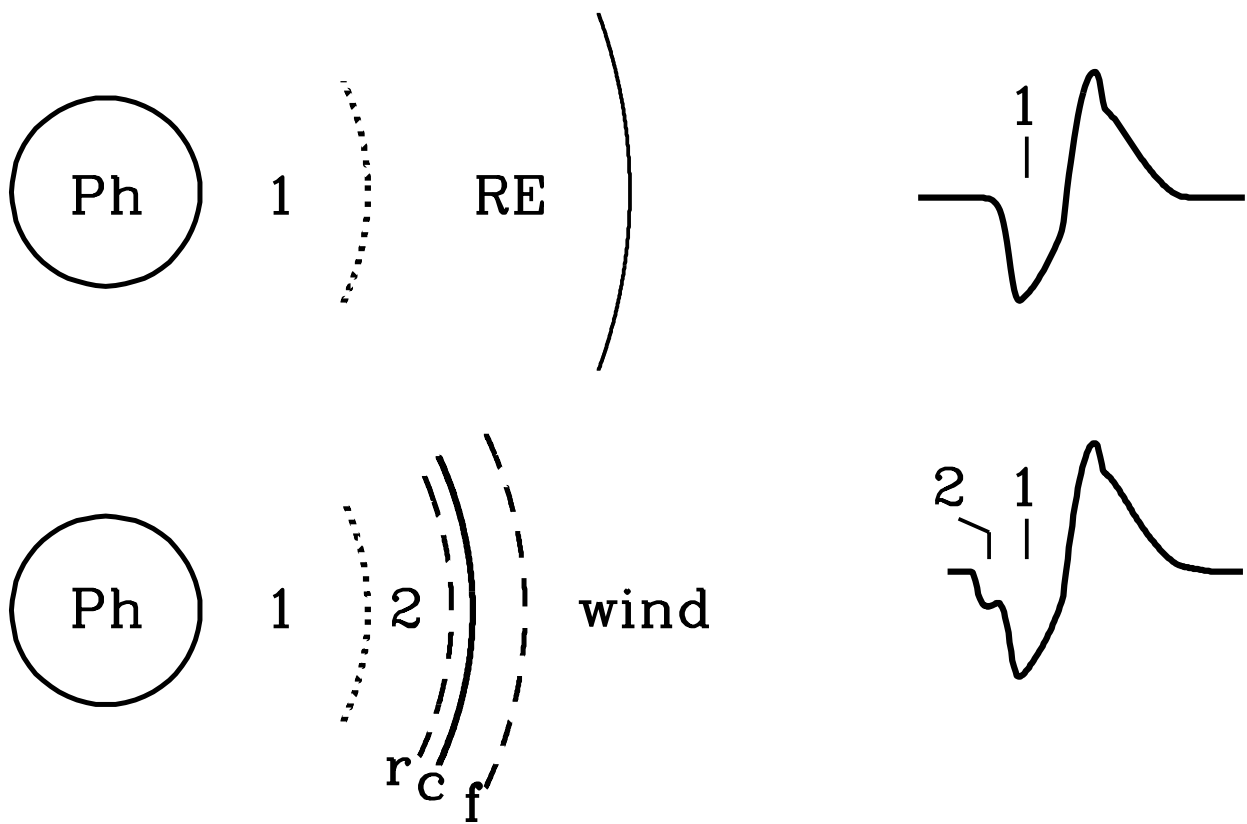}
  \caption{Schematic picture of the formation of H$\alpha$ 
without and with CS interaction.
In the absence of CS interaction (upper), the absorption component
forms in the inner layers of ejecta ({\em 1}) against the photosphere 
({\em Ph}); the outer recombined ejecta ({\em RE}) do not 
contribute to the absorption
line profile (upper right). With CS interaction (lower), the double-shocked 
structure arises at the SN/wind interface with the forward shock ({\em f}),
reverse shock ({\em r}) and contact surface where the cool dense shell occurs 
 ({\em c}). The X-rays, primarily from the reverse shock, ionize and 
excite the 
outer layer of ejecta ({\em 2}) where the HV absorption forms, producing 
a depression 
in the blue wing of the undisturbed absorption (bottom right).
}
   \label{f-cart1}
  \end{figure}

The model line profiles of H$\alpha$ and He\,I 10830 \AA\
are computed for different wind densities,
ages, ejecta mass and energy (Table 1).
The standard model is characterized by $M=19~M_{\odot}$,
$E=1.3\times10^{51}$ erg, $k=9$. In all cases
 $v_{\rm c}=15,000$ km s$^{-1}$ with  
the exception of the very low energy model (eW) for which 
$v_{\rm c}=6500$ km s$^{-1}$.
We designate a model at a particular time by adding the
age in days to the model name; e.g., W50 stands for model W on day 50.
The effect of the wind density on H$\alpha$ on day 50
is shown in Fig. \ref{f-modha}{\em a} for the wind density
parameters  $w=0.5$, 1, 2, and 4. Without a wind,
the H$\alpha$ absorption on day 50 forms because of the time-dependent
ionization effect; a high ionization is maintained due to Ly$\alpha$
trapping and reionization from excited levels of hydrogen. 
The CS interaction results in the ionization and 
excitation of the outer recombined layers of 
unshocked ejecta and, as a result,
a HV absorption feature appears in the blue wing of the undisturbed profile.
Fig. \ref{f-cart1} illustrates how H$\alpha$ absorption 
is modified by the CS interaction.
For $w=0.5$, the CS interaction effect is small (Fig. \ref{f-modha}{\em a}) 
and the wind density $w=0.5$ seems to be the least that could be detected
by HV absorption in H$\alpha$. For $w\sim1$, the strength of 
the HV absorption increases with
the wind density roughly as $w^2$ and for $w>2$ becomes saturated.
The decrease of the velocity of the blue absorption edge is
due to the deceleration caused by the dense wind. 
The fact that for $w=4$ the HV absorption merges with undisturbed 
H$\alpha$ absorption implies that in the case of a very dense wind 
it would be difficult to infer the density parameter $w$ 
using only the H$\alpha$ absorption.

The evolution of the HV absorption (Fig. \ref{f-modha}{\em b}) shows
that there is an optimal phase at about the middle of the
plateau ($\sim50$ d) when the CS interaction effect is most
pronounced. At an early stage (e.g., on day 20) the
CS interaction effect is not clearly seen because of the
 merging of HV absorption with the  strong undisturbed absorption, while at a
late stage (e.g., 80 d) HV absorption becomes very faint.
The sensitivity of HV absorption to the ejecta density is shown 
by the model mW50 with a mass $10~M_{\odot}$ and energy 
$9\times10^{50}$ erg.  The energy is chosen to 
match the blue wings of the main absorption of models mW50 and W50. 
The HV absorption in the low mass case turns out the same as in the model W50, 
i.e., the effect of reducing the ejecta density by a factor of 2 is negligible.
We also consider the CS interaction effect for SNe~IIP with a very low
kinetic energy, $E=3\times10^{50}$ erg, which presumably applies to the
low-luminosity SNe~IIP, e.g., SN~1999br \citep{Pas04}. This model 
with $v_{\rm c}=6500$ km s$^{-1}$ and $w=1$ 
predicts strong HV absorption which is merged with the 
undisturbed absorption (Fig. \ref{f-modha}{\em d}).
For all the models when HV absorption is unsaturated, the Sobolev 
optical depth of this feature for $w\sim 1$ is $\tau\leq1$. This implies 
that in the H$\beta$ line
the optical depth $\leq0.14$, so we do not expect
that HV absorption in this line will be pronounced
 (but see \S~\ref{sec-narrow}).

We examined the sensitivity of HV absorption to the power law index $q$ of the
X-ray spectrum of the reverse shock. We found that a softer
spectrum ($q=1.2$) and
harder spectrum ($q=0.8$) produce slightly deeper HV absorption, so
the preferred case $q=1$ corresponds approximately to a minimal
effect of the CS interaction on H$\alpha$. We checked
the sensitivity of HV absorption to the adopted SN bolometric luminosity as well.
The HV absorption strength anticorrelates with the
bolometric luminosity, but the dependence is weak. The mechanism
for this dependence is the
depopulation of excited levels by photoionization.

\begin{figure}
\plotone{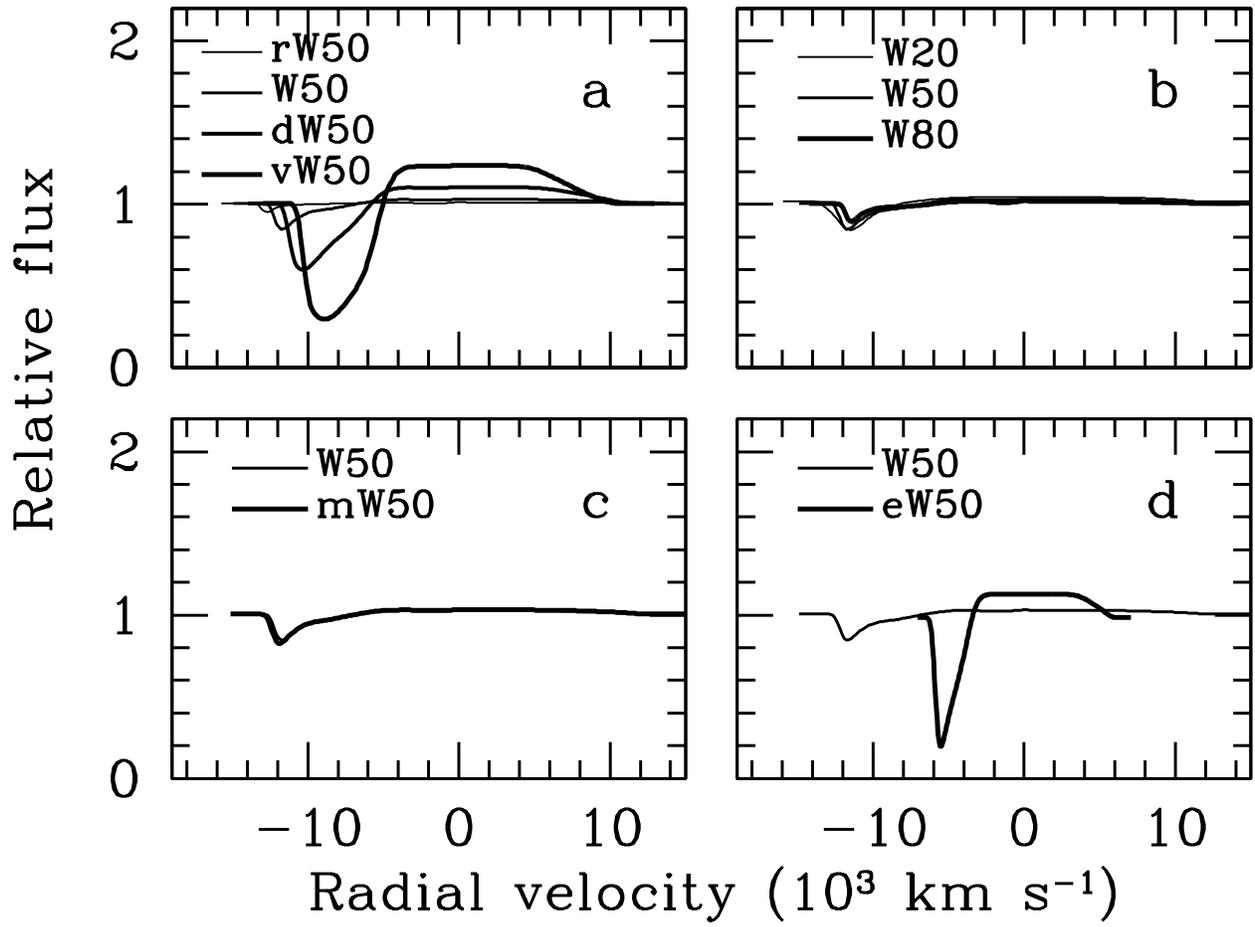}
  \caption{The same as Fig. 2 but for the He\,I 10830 \AA\ line.
}
   \label{f-modhe}
  \end{figure}

The CS interaction effect in the He\,I 10830 \AA\ line substantially
differs from that in H$\alpha$ in one important respect: in the
absence of a wind
the model does not predict noticeable absorption in He\,I 10830 \AA\
(Fig. \ref{f-modhe}{\em a}), except for a very early stage ($t<10$ d).
The wind density $w=0.5$ produces weak
HV absorption, on the verge of detectability, while the case $w=1$ yields
HV absorption with a
relative depth of $\sim0.2$ and could be easily detected.
The absorption gets markedly stronger for larger $w$, showing
the sensitivity of He\,I absorption to the wind density in this
particular range  of the $w$ parameter.
Remarkably, while in H$\alpha$ the contribution of the interaction 
effect for $w=4$ is difficult to discern (Fig. \ref{f-modha}{\em a}),  
 the He\,I 10830 \AA\ 
HV absorption provides a probe for such a dense wind.

The strength of HV absorption in He\,I 10830 \AA\ increases between days 20 and 50
but gets weak on day 80 (Fig. \ref{f-modhe}{\em b}), resembling
the behavior of HV absorption in H$\alpha$.
The effects of the ejecta density (Fig. \ref{f-modhe}{\em c}) 
 are similar to those in H$\alpha$.
The HV absorption for the low energy case is very strong and easily discernable,
 unlike in H$\alpha$.


\subsection{Comparison with observations}\label{sec-observ}


\subsubsection{HV absorption in H$\alpha$}

\begin{figure}
\plotone{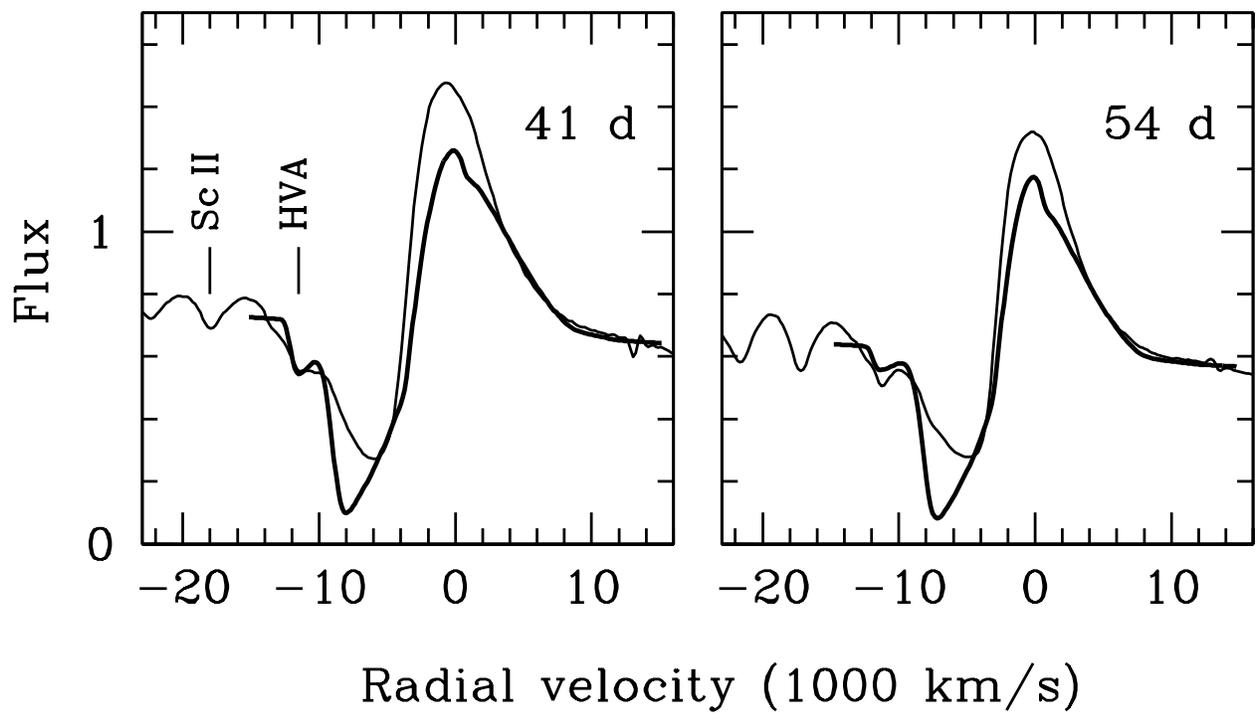}
  \caption{Modeling of the H$\alpha$ profile with  HV absorption
 in SN~1999em on days 41 and 54.
The model ({\em thick} line) is overplotted on the observations
  (Elmhamdi et al. 2003).
}
  \label{f-ha99em}
  \end{figure}

Spectra of SN~1999em \citep{Elm03}  on days 41 and 54 (assuming the
 explosion date JD 2451476) show a depression in the blue wing
of the H$\alpha$ line (Fig. \ref{f-ha99em}) at the predicted
position of HV absorption ($-10,000$ to $-12,000$ km s$^{-1}$).
To model the H$\alpha$ profiles we adopted the standard parameters
$M=19~M_{\odot}$, $E=1.3\times10^{51}$ erg, $k=9$, and
$v_{\rm c}=15,000$ km s$^{-1}$. We find the best value of
the wind density parameter is $w=1$ with an uncertainty of
$\sim\pm0.07$, assuming the other parameters to be fixed.
The indicated error reflects only the accuracy of estimating this 
parameter from the best fit and does not include errors in 
the observed spectrum and uncertainties in the model. 
Also on day 54, the model reproduces the general strength of
the observed HV absorption (Fig. \ref{f-ha99em}). Yet we 
find that the observed HV absorption on day 54 seems to 
contain a notch component that is deeper than in the model.

The question arises of why the model fails to describe in detail the undisturbed H$\alpha$.  
The model emission is weaker, while the absorption is stronger, than is
observed \citep[Fig. \ref{f-ha99em} and][]{U07}. The weak 
model emission is probably related to
our crude description of the UV continuum in the SN atmosphere
\citep[cf.][]{UC05}.
However, this cannot explain why the observed absorption is shallow.
In fact, a check of available spectra of SNe~IIP 
indicates that the H$\alpha$ absorption seems to be shallow
in any normal SN~IIP (e.g., SN~2004dj, see below).
In contrast, for the peculiar SN~1987A related to the explosion
of a blue supergiant, the H$\alpha$ absorption is
deep and  well reproduced by the time-dependent model \citep{UC05}.
We speculate that this is a strong indication that the shallowness of
the H$\alpha$ absorption in normal SNe~IIP is 
related to the red supergiant structure of the presupernova.
For instance, convection in the red supergiant atmosphere
could produce large amplitude density perturbations
that are significantly amplified during the blast wave propagation.
As a result, the outer layers of ejecta would acquire a clumpy structure
responsible for the shallowness of the H$\alpha$ absorption. 

\begin{figure}
\plotone{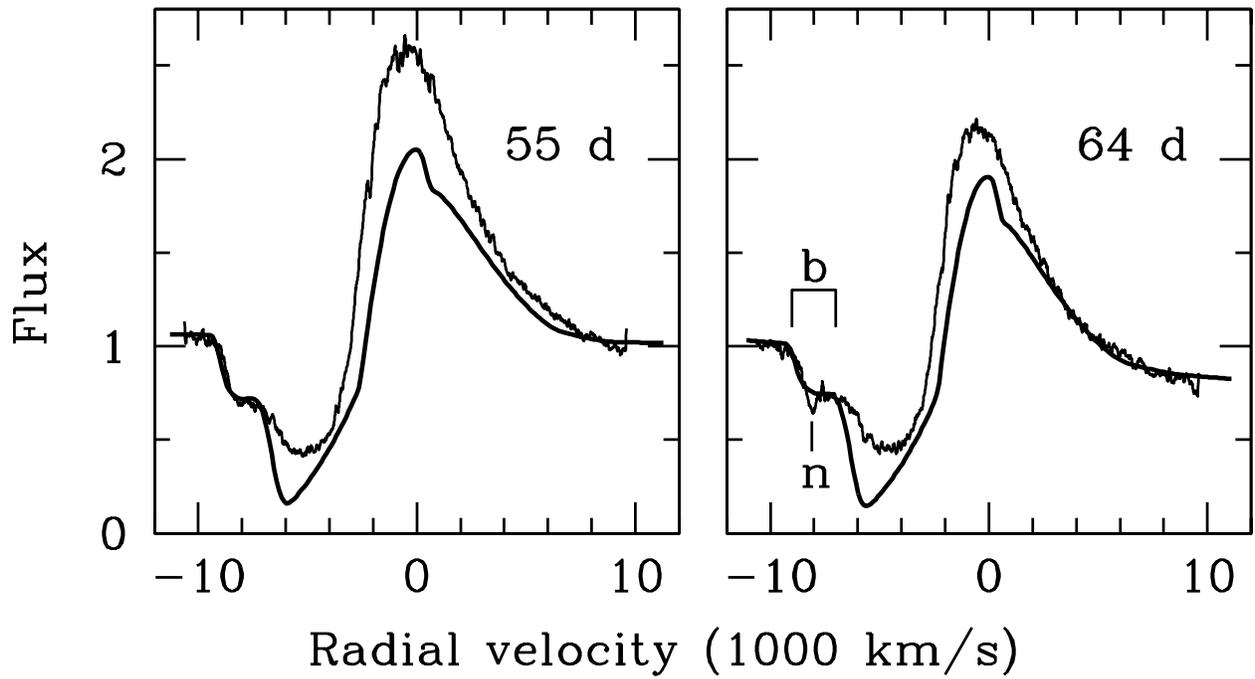}
  \caption{Modeling of the H$\alpha$ profile with HV absorption 
in SN~2004dj on days 55 and
64. The model ({\em thick} line) is overplotted on the observations
  (Korcakova et al. 2005). On day 64, a notch (n) appears 
against broader HV absorption (b).
}
  \label{f-ha04dj}
  \end{figure}

Returning to the HV absorption, in Fig. \ref{f-ha04dj}
we show the model fit to the observed H$\alpha$ in SN~2004dj 
\citep{Kor05} on days 55 and 64 (for an explosion date JD 2453185).
The model parameter set is $M=19~M_{\odot}$, $k=9$, as for SN~1999em, but
$E=6.7\times10^{50}$ erg.
One could retain
the energy $1.3\times10^{51}$ erg used for SN~1999em 
and vary $M$. However, in this case one requires
$M\approx50~M_{\odot}$ to attain a comparable fit.
Since the ejecta mass unlikely exceeds $20~M_{\odot}$, the lower
energy case for SN~2004dj is preferred.
In accord with the lower energy we take 
$v_{\rm c}=13,000$ km s$^{-1}$ and 
lower velocities at the photosphere by a factor 0.9 compared to SN~1999em.
The derived wind density
($w=1.27\pm0.1$) is determined by the strength of the HV absorption and its position,
while the lower energy is needed to reproduce the correct position of
the HV absorption. To check the effect of the adopted boundary velocity 
$v_{\rm c}$, we modeled the case $v_{\rm c}=11,000$ km s$^{-1}$. 
In this case the wind density is lower, $w=0.9$. We prefer 
$v_{\rm c}=13,000$ km s$^{-1}$ on the basis of the arguments presented below 
in Sec. \ref{sec-conduct}. 

On day 55 the HV absorption in the observed spectrum has the appearance of a
relatively broad shallow depression similar to SN~1999em on day 41. A bit later,
on day 64,  a notch (``V-shaped'' narrow absorption) 
appears at a radial velocity $\sim-8200$ km s$^{-1}$  (Fig. \ref{f-ha04dj}).
The feature dominates the HV absorption 
until at least day 102 in spectra of SN~2004dj reported by \cite{Vin06}.
Neither the strength nor the shape of the notch at the late photospheric 
stage $t\geq60$ d can be produced by 
a model in which the HV absorption forms only in the unshocked ejecta.
This suggests that the notch in H$\alpha$ has a 
{\em different origin} which is addressed in \S~\ref{sec-narrow}.

The HV absorption in SN~2004dj has a markedly lower velocity
compared to SN~1999em at a similar epoch (8200 vs.\ 11,500 km s$^{-1}$).
This fact is significant for the origin of the depression in the blue wing
of H$\alpha$ since it rules out the possibility that the feature might be
produced by some unidentified metal line.
Photospheric velocities at this epoch are similar
in both SNe within several $100$ km s$^{-1}$,
so the expected positions of a metal line absorption
should coincide in both SNe within the same range of velocities.
Since this is not the case, we conclude that HV absorption
in these supernovae {\em cannot be} produced by an unidentified line.
An additional argument against the `metal line' option is
that SN~1987A does not show  HV absorption in H$\alpha$ at all.
This would be difficult to understand in the context of the
metal line option, but is easily explained by the very low density
of the pre-SN wind \citep[$w\sim 2\times10^{-3}$,][]{CD95} 
that is incapable of producing detectable interaction effects.

\subsubsection{HV absorption in He\,I 10830 \AA\ }

\begin{figure}
\plotone{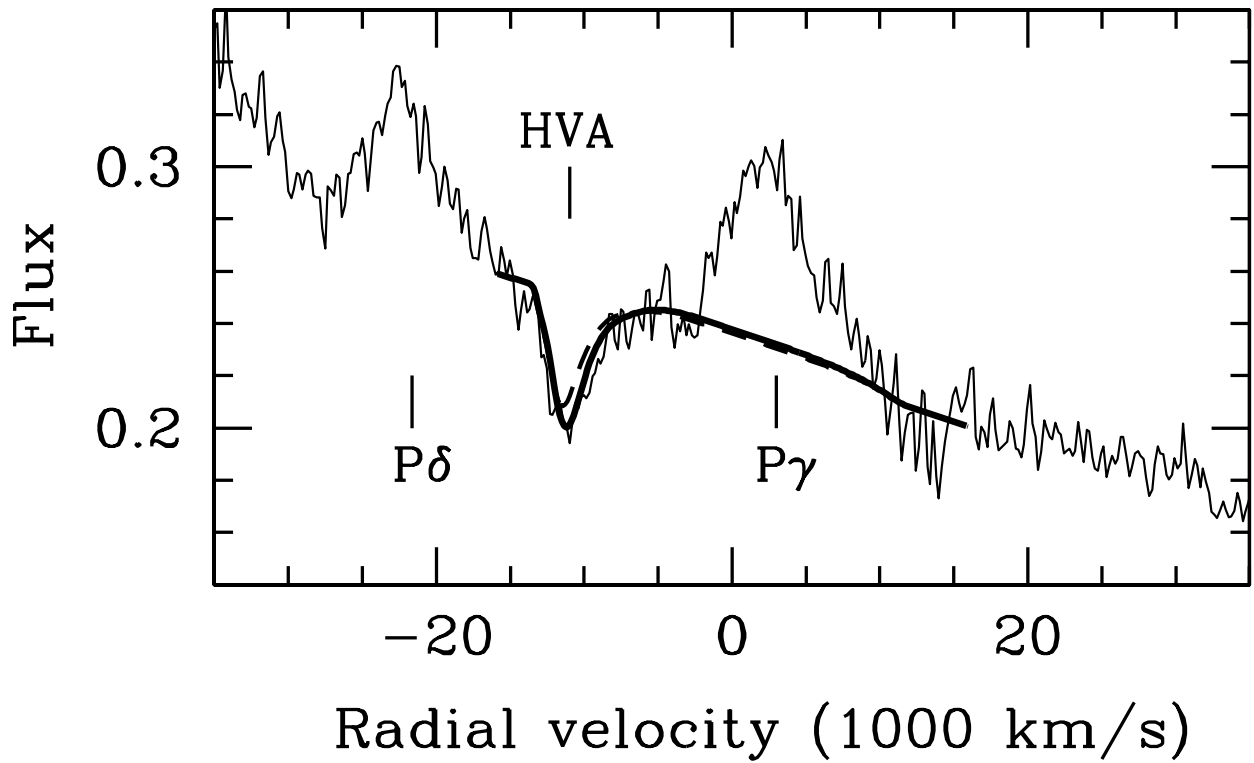}
  \caption{ High velocity absorption in the He\,I 10830 \AA\ line in
SN~1999em on day 20.
The model is overplotted on the observations
  (Elmhamdi et al. 2003).   The {\em thick} line is the model for
$w=1.15$ and the {\em dashed} line is the model with $w=1$.
Vertical bars show the rest positions of
  Paschen emission lines.}
  \label{f-he99em}
  \end{figure}

The HV absorption of He\,I 10830 \AA\ in SN~1999em on day 20
was identified by \cite{Elm03} and
attributed originally to radioactive excitation by external $^{56}$Ni.
Here we abandon this explanation and propose excitation of He in the
external layers by CS interaction.
The He\,I 10830 \AA\ line of SN~1999em on day 20  \citep{Elm03}
is shown in Fig. \ref{f-he99em} together with
models for $w=1$ and the best value $w=1.15$, which has an uncertainty of
$\pm0.07$.
The wind density parameter derived from the HV absorption strength in He\,I 10830 \AA\
agrees well with the value $w=1$ found from the H$\alpha$ line.
The clear-cut observed profile and the sensitivity of HV absorption of
He\,I 10830 \AA\ to the wind density
makes this line a valuable diagnostic tool for probing the CS wind in SNe~IIP.

Interestingly, based on a time-dependent model for a SN~IIP spectrum without 
CS interaction, \cite{D07} predict strong HV absorption 
in He\,I 10830 \AA\ at a relatively late epoch; 
on day 50, the relative depth of He\,I 10830 \AA\ 
is $\approx 0.1$ for an enhanced He abundance 
$n(\rm{He})/n(\rm{He})=0.2$. This suggests a relative depth of $\approx0.05$ 
for a solar He abundance. The absorption lies 
in the range of  $-12,000$ to $-17,000$ km s$^{-1}$, which corresponds 
to the blue wing of He\,I 10830 \AA\ in SN 1999em on day 20 (Fig. \ref{f-he99em}). 
The He\,I 10830 \AA\ line in SN~1999em on day 20 does
not show  evidence for the additional He absorption in the blue wing. 
Yet it is possible that in a more detailed model the
time-dependent effects for He lines might be larger than in our simple model 
and would substantially contribute to the HV absorption of 
the He\,I 10830 \AA\ line at an age $t>20$ d. However, for SN~1999em on 
day 20, we note the close correspondence between  the
$w$ values deduced from the H$\alpha$ and He\,I 10830 \AA\ lines.

Spectra of SN~1999em covering the He\,I 10830 \AA\ line are also reported 
by \cite{Ham01} for three epochs: 8, 24, and 34 d. The first
spectrum shows a strong He\,I 10830 \AA\ line with a P Cygni profile.
This line, related to the early high temperature phase, 
rapidly disappears and is not seen in the next two spectra.
Instead, HV absorption appears in this line.
On day 24, the HV absorption  of He\,I 10830 \AA\ is much the same as 
on day 20 in Fig. \ref{f-he99em}, and it also appears on day 34.
A steady HV absorption feature with time is expected in our
model (\ref{f-modhe}b).
On day 34, an additional redder absorption at 
$\sim10549$ \AA\ appears separate
from the HV absorption. The transformation of the He\,I 10830 \AA\ line
 from a single HV absorption into a double absorption structure 
is also observed between days 22 and 44 in another Type IIP supernova,
 SN~1995V \citep{Fa98}.
We believe that the low velocity absorption is the result of He excitation by
gamma-ray photons leaking from the inner radioactive
decays of $^{56}$Ni and $^{56}$Co \citep{Fa98,Chu87}. To produce the required
excitation of He about $\sim10^{-5}~M_{\odot}$ of $^{56}$Ni should be 
mixed out to a velocity $\sim 4000$ km s$^{-1}$, which 
does not contradict   the conclusion that the 
majority of the $^{56}$Ni resides in deeper layers with velocity
 $\leq600$ km s$^{-1}$ 
\citep{U07}. 


\section{ORIGIN OF THE NOTCH IN H$\alpha$}\label{sec-narrow}

We now address the issue of the notch that emerged in
H$\alpha$ in both SN~1999em and SN~2004dj at about day $60$.
We propose that the notch is produced by 
H$\alpha$ absorption of the photospheric radiation in the CDS. 
We also consider an alternative mechanism:
additional excitation of hydrogen in the
unshocked ejecta by Ly$\alpha$ photons emitted from the CDS.


\subsection{CDS vs. unshocked ejecta}\label{sec-alternative}

\begin{figure}
\plotone{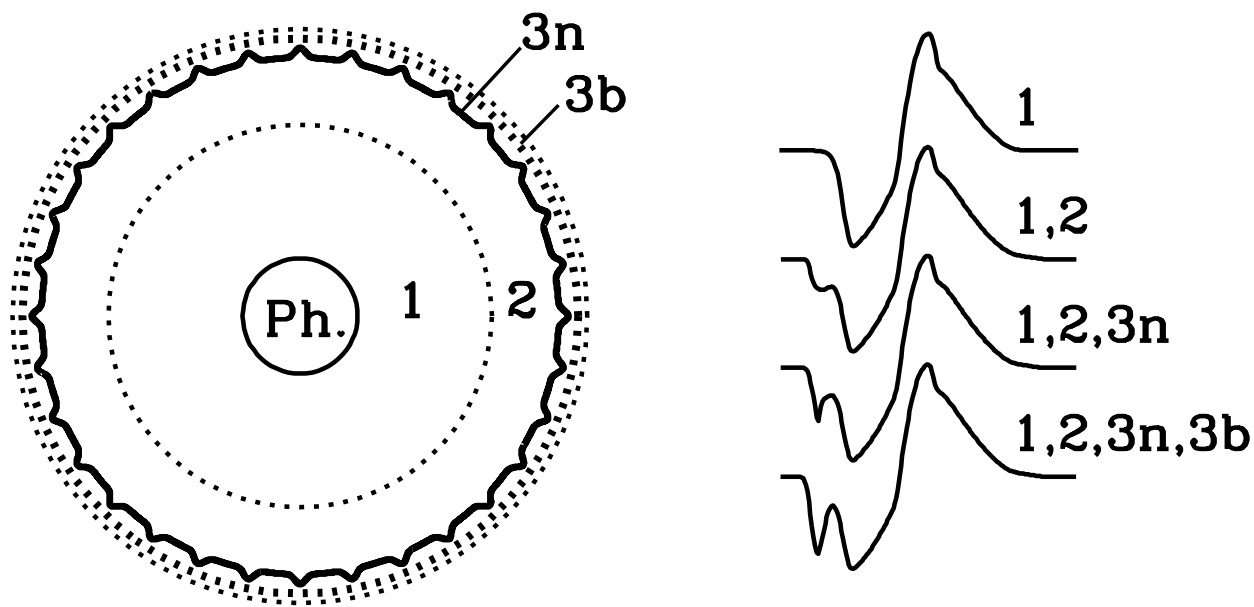}
  \caption{Visualization of   H$\alpha$ formation with a 
contribution from the CDS.
{\em Left}: The cartoon of a SN depicts the photosphere ({\em Ph.}),
the undisturbed atmosphere responsible for  
H$\alpha$ in the absence of CS interaction
({\em 1}), the layer excited by X-rays from the reverse shock ({\em 2}) responsible for 
the shallow ejecta HV absorption, the CDS ({\em 3n}) responsible for the narrow notch,
and
the CDS gas mixed into the forward shock 
region ({\em 3b}) responsible for the broad notch.
{\em Right, upper to lower}: The H$\alpha$ profile without CS interaction ({\em 1}),
 with the contribution of the ejecta HV absorption ({\em 1,2}), with the 
contribution of a narrow CDS component ({\em 1,2,3n}), and with the added 
broad CDS component ({\em 1,2,3n,3b}).
}
  \label{f-cart}
  \end{figure}

We first study the preferred explanation for the notch, 
 H$\alpha$ absorption in the CDS, and consider
the clearly observed case of SN~2004dj.
In modeling the H$\alpha$ absorption produced by the CDS, the 
principal parameters are the column density of the cool gas, 
and its excitation temperature and turbulent velocity. 
The undisturbed or weakly disturbed CDS is characterized by 
a turbulent velocity of the order of the thermal velocity of the 
cool gas,  $\sim 10-20$ km s$^{-1}$. However, with this turbulent 
velocity the equivalent width of the HV absorption turns out to be
too small. A possible solution is to increase 
the turbulent velocity, as expected by the action of the 
Rayleigh-Taylor (RT) instability of the decelerating thin shell.
A significant deceleration of the model CDS of SN~2004dj,
about 20\% during the first 10 days (see low energy 
case in Fig. \ref{f-dyn}), favors 
the development of the RT instability of the CDS followed by mixing
of the cool dense gas with hot forward shock gas \citep{CB95,BE01}.
The mixed cool gas thus can acquire a significant velocity dispersion and
turbulent velocities up to $\sim 0.1$ of the shell velocity.
Both components could contribute to the notch formation:
absorption in the weakly disturbed CDS and in the mixed CDS material.
We therefore consider two components: (1) a weakly 
perturbed spherical shell with a
 turbulent velocity of $u_{\rm t}\approx20$ km s$^{-1}$ and (2) mixed
CDS gas with a large turbulent velocity $u_{\rm t}\approx400-500$ km s$^{-1}$.
The mass fraction of the second component is denoted $\beta$.
The corresponding spectral components of the HV absorption related to the CDS 
can be dubbed 
 narrow notch and broad notch. These components should be distinguished from 
the `ejecta component' -- a broader shallow HV absorption formed in the unshocked ejecta and addressed in the previous section.
A cartoon (Fig. \ref{f-cart}) illustrates the origin the HV absorption 
produced by the ejecta and the CDS.

Hydrogen excitation of the CDS gas is presumably
maintained by the absorption of X-rays emitted from the
reverse shock. The absorbed energy of  X-rays is assumed to be
homogeneously distributed through the CDS material, as expected
for the modest shell column densities for our situation.
The number density in the CDS is determined by pressure equilibrium
$P\approx \rho_{\rm w}v_{\rm s}^2$ (where $\rho_{\rm w}$ is the pre-shock wind 
density) for the typical temperature of the cool gas of $\sim10^4$ K 
implied by the energy balance. 
On day 64, the CDS mass in our model for SN~2004dj
is $3.2\times10^{-4}~M_{\odot}$. The other input parameters provided by the 
interaction model are:
 CDS radius, $R_{\rm s}=5.2\times10^{15}$ cm, velocity
$v_{\rm s}=8350$ km s$^{-1}$, X-ray luminosity of the reverse shock
$L_{\rm X}=1.53\times10^{38}$ erg s$^{-1}$, and electron temperature in the
reverse shock $kT_{\rm X}=1.36$ keV.
The population of the second level of hydrogen is computed
using a steady-state approximation (justified since $n_e>10^8$ cm$^{-3}$),
 and assuming a two level plus continuum model for the hydrogen atom.
Both non-thermal excitation and ionization by secondary electrons
lead to the excitation of the second level, while depopulation
 proceeds by the two-photon transition, Ly$\alpha$ escape and collisional
deexcitation.

\begin{figure}
\plotone{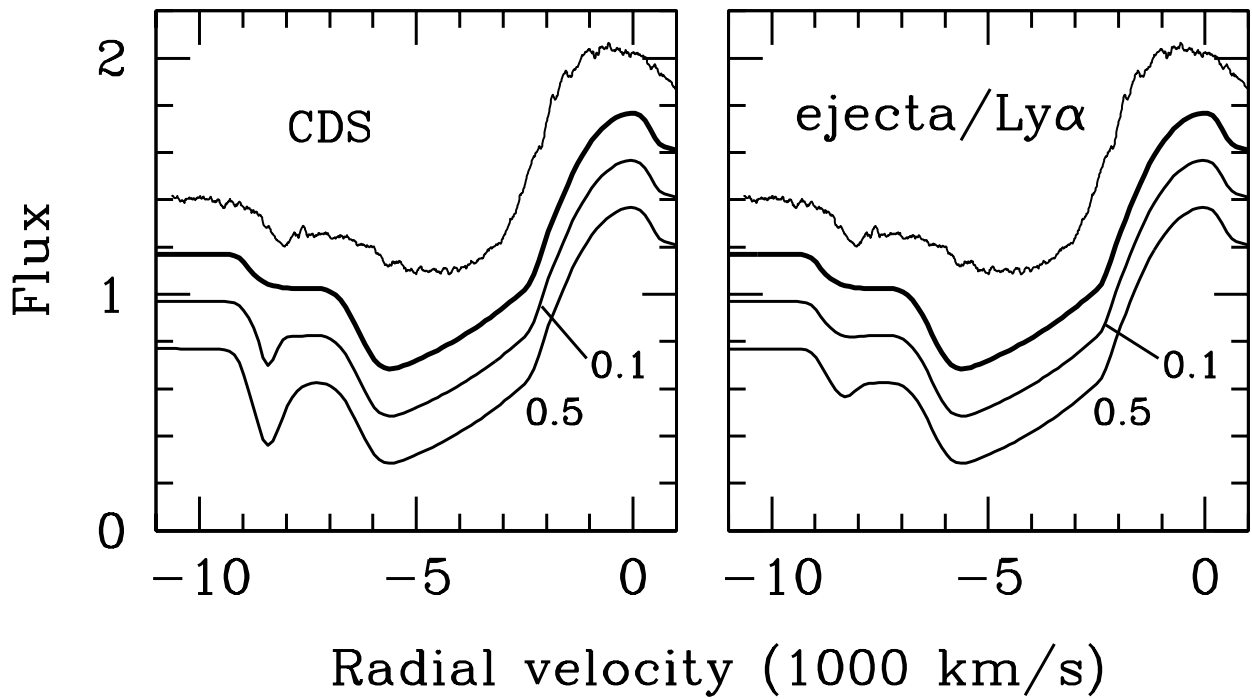}
  \caption{Modeling of the HV absorption of H$\alpha$ in SN~2004dj on day 64.
The {\em left} panel shows, from top to bottom, the observed spectrum, 
a model without
the CDS absorption, and models with the notch formed in the CDS 
assuming a two-component CDS with turbulent velocities of
20 and 500 km s$^{-1}$ for different fractions of broad components
(0.1 and 0.5).
On the {\em right} panel, the two top lines are the same as on the left,
while the next two are models of the notch formed in the
unshocked ejecta by Ly$\alpha$ absorption.
The curves are labelled by the corresponding efficiency
 of Ly$\alpha$ emission.
}
  \label{f-nHVA}
  \end{figure}

The calculated profiles of HV absorption related to the CDS 
are overplotted on the model H$\alpha$ profile with ejecta HV absorption
and displayed in Fig. \ref{f-nHVA} along with the
observed spectrum of SN~2004dj \citep{Kor05} on day 64.
Shown are three cases: one without the CDS absorption and two
cases with the CDS absorption
that differ by the mass fraction of the broad CDS component
($\beta=0.1$ and 0.5). For the quoted values, the
optical depths of the narrow components are 520 and 175,
while these values for the broad component are 0.2 and 0.45,
respectively.

The absolute value 
of the radial velocity of the HV absorption produced by the 
CDS ($v_{\rm a}$)
is slightly
lower than the velocity of the CDS, $v_{\rm s}$, owing to the sphericity effect.
Assuming uniform brightness of the photosphere (Lambert's law)
it is straightforward to show that the average absolute value of the
radial velocity of the absorption is
\begin{equation}
v_{\rm a}=\frac{2}{3}\frac{v_{\rm s}}{x^2}\left[1-(1-x^2)^{3/2}\right]\,,
\label{eq-hva}
\end{equation}
where $x=v_{\rm p}/v_{\rm s}$ is the ratio of the photospheric to the CDS
velocity. According to this expression $v_{\rm a}<v_{\rm s}$ and,
in the limit $x\ll1$,  $v_{\rm a}=v_{\rm s}$ with high accuracy.
The sphericity effects are responsible also for the additional
broadening of the absorption over the radial velocity range
$(v_{\rm s}^2-v_{\rm p}^2)^{1/2}<|v_r|< v_{\rm s}$, which is
$\sim460$ km s$^{-1}$ for the model of SN~2004dj on day 64.

The alternative mechanism is that the notch could
form as a result of an enhancement of hydrogen excitation in the unshocked
SN ejecta due to the scattering of Ly$\alpha$ photons
emitted from the CDS that in turn is irradiated by X-rays from the 
reverse shock. A resonance condition implies that
Ly$\alpha$ photons can scatter in the unshocked ejecta only in
the narrow velocity range $v_{\rm s}<v<v_{\rm sn}$, where
$v_{\rm sn}$ is the boundary velocity of the unshocked SN ejecta.
The expected flux of Ly$\alpha$ from the CDS is
\begin{equation}
F_{12}=\frac{1}{2}\psi\frac{L_{\rm X,abs}}{4\pi R_{\rm s}^2}\,,
\end{equation}
where $\psi$ is the efficiency of Ly$\alpha$ emission and $L_{\rm X,abs}$ is
the X-ray luminosity absorbed by the CDS.
The average excitation rate in the ejecta
within the velocity width $\Delta v=v_{\rm sn}-v_{\rm s}$
is balanced by Ly$\alpha$ escape; other
depopulation mechanisms are negligible. The
population of the second level ($n_2$) thus is approximately determined
by the rate equation
\begin{equation}
n_2A_{21}\beta_{12}=\frac{F_{12}}{h\nu_{12}\Delta vt}\,,
\end{equation}
where $t$ is the SN age, $h\nu_{12}$ is the Ly$\alpha$ photon energy, 
and $\beta_{12}$ is the Sobolev escape probability for Ly$\alpha$.
For the model of SN~2004dj,
the notch in H$\alpha$ produced by Ly$\alpha$
scattering in the ejecta is calculated on day 64 (Fig. \ref{f-nHVA}).
Two cases, with efficiencies $\psi=0.1$ and 0.5, are plotted. One sees
that even for
the unacceptably high efficiency  $\psi=0.5$ the notch 
produced by the Ly$\alpha$ scattering
is significantly weaker than that predicted by
the absorption in the CDS. We conclude that the latter mechanism
is the most likely explanation of the notch in H$\alpha$.

We examined whether the absorption in the CDS might produce
some effect in the He\,I 10830 \AA\ line. For the above model,
the optical depth in the He\,I 10830 \AA\ line turns out to be
a factor of $\sim10^2$ lower than in H$\alpha$.
Therefore, it is unlikely that the notch can be seen in He\,I 10830 \AA.
The main reason is the efficient depopulation of the 2$^3$S level 
in the CDS by Penning ionization of hydrogen.

\begin{figure}
\plotone{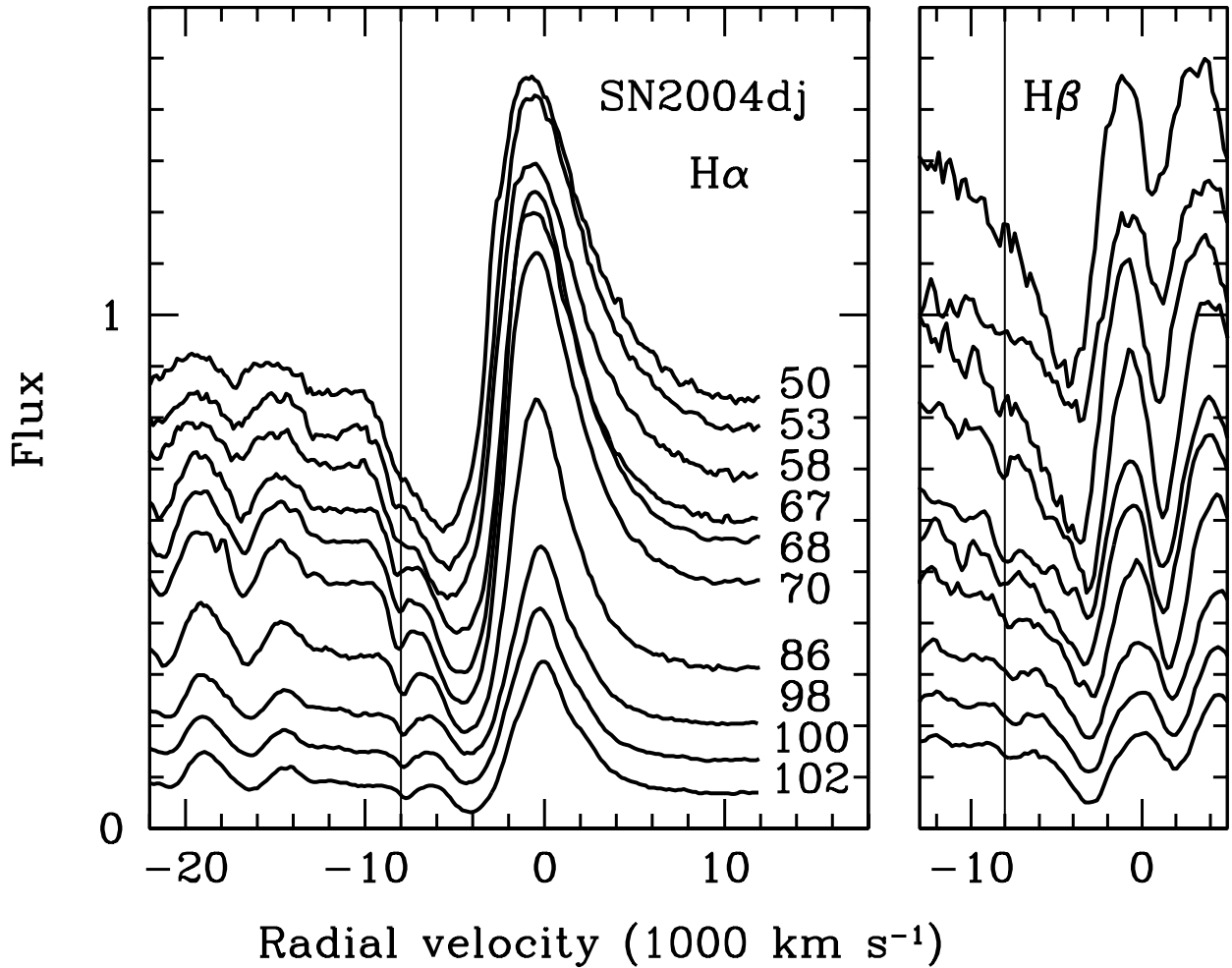}
  \caption{A sequence of SN~2004dj spectra \citep{Vin06}
 in the regions of the H$\alpha$ and H$\beta$ lines for the same epochs, showing
the HV absorption.
Next to the H$\alpha$ spectra are shown the corresponding ages; 
the ages of the H$\beta$
spectra are the same. The notch is seen in both lines at a radial
velocity of about $-8000$ km s$^{-1}$ starting on day 67.
}
  \label{f-obs04dj}
  \end{figure}

Given the relatively high H$\alpha$ optical depth of the CDS,
a notch at the same velocity could be seen in H$\beta$ as well.
We inspected spectra of SN~1999em \citep{Leo02} and found the signature of
a notch in H$\beta$ at the right position ($\approx-11,500$ km s$^{-1}$)
between days 39 and 81. In fact, \cite{Leo02} already discussed the
presence of the HV absorption in H$\alpha$ and H$\beta$ in  spectra of SN~1999em
and concluded that these features cannot be caused by metal line absorptions.
We also find a notch in H$\beta$ in  spectra of SN~2004dj
obtained by \cite{Vin06}.
Since the HV absorption components of H$\alpha$ and H$\beta$
have not been previously mentioned for this
supernova, we show parts of the SN~2004dj spectra \citep{Vin06}
retrieved from the SN archive SUSPECT (Fig. \ref{f-obs04dj}).
The spectra show conspicuous notches at
about $-8000$ km s$^{-1}$ both in H$\alpha$ and H$\beta$ between days 67 and 102.
Below we will model HV absorption in both lines in more detail.


\subsection{Rayleigh-Taylor mixing effects in HV absorption}\label{sec-conduct}

Here we present a model that can account for the strength and shape of 
the notch in H$\alpha$ of SN~2004dj at the late photospheric epoch.
Preliminary modeling suggests that the simple two-component model of the 
CDS (narrow and broad) introduced above is not able to 
describe the strength of the notch at late times ($\sim100$ d): 
the excitation produced by the X-ray absorption is insufficient. 
This suggests the presence of a more efficient excitation 
mechanism or an additional source of heat.
Another problem is the relative depth of notches in 
the H$\alpha$ and H$\beta$ lines: they are  
comparable and shallow. This indicates that some veiling 
is present, possibly due to clumpiness. Both reasons force us 
to modify the two-component model of the CDS absorption. 

An important outcome of the RT mixing of the CDS material 
in the forward shock is the growth of the
area of the contact surface ($S$) between cool and hot gas.
The development of mixing may be thought as the progressive decrease
of the minimum linear scale of fragments ($\lambda$).
This process is demonstrated in numerous experiments
on the RT instability and in the case of SN/CS interaction is clearly
seen in 2D and 3D hydrodynamical computations \citep{BE01}.
A fragmentation cascade of a fixed amount of cool gas
leads to the scaling $S\propto 1/\lambda$, while for a stationary
mixing process $S\propto \lambda^{-1/3}$ \citep{SRM89}.
The growth of $S$ and, accordingly, of the surface-to-volume ratio for the
mixed cool gas, should lead to a greater role of thermal conduction
in the heating of the cool gas at late times. 
We speculate that this
is the missing factor that could resolve the excitation problem.

To describe the effect of thermal conduction, we introduce a third CDS component: an additional broad CDS component powered predominantly 
by thermal conduction. The mass fraction of the third 
component is $\mu\ll1$, while the 
cumulative area of the contact surface for the third
component, $S=C_{\rm s}4\pi R_{\rm s}^2$, can be large, with an
area ratio $C_{\rm s}\gg1$. We assume that the third component
is homogeneously heated by thermal conduction,
which is a sensible approximation for a small thickness of sheets of CDS
gas, $\sim10^9$ cm,
comparable to the mean free pass length of hot electrons and protons in the
cool dense gas. The heat flux is presumably a small
fraction of the saturated flux $F_{\rm c}=\eta_{\rm c}F_{\rm sat}$;
we adopt $\eta_{\rm c}=0.01$ to
allow qualitatively for the possible suppression of thermal conduction by
a magnetic field. The saturated flux is taken of the form
$F_{\rm sat}=0.4(2kT_e/\pi m_e)^{1/2}n_ekT_e$ \citep{CM77}.
The forward shock is not equilibrated ($T_e\ll T_i$), so
the ion (proton) conduction could be significant.
In our case, the ratio of proton-to-electron thermal flux is
$(T_i/T_e)^{3/2}(m_e/m_p)^{1/2}\sim2$ and this fact is taken into account.
Ionization and thermal balance are solved for the dense hydrogen
cooled by Ly$\alpha$ and two-photon radiation.

The RT mixing generally produces  
an inhomogeneous (lumpy) distribution of mixing regions.
This is the result of
the existence of a dominant angular scale of the RT
 instability, about $10^{\circ}-15^{\circ}$ \citep{CB95,BE01} and
of the intermittence of the turbulent mixing.  
We describe the distribution of mixed regions as
an ensemble of isolated lumps which occupy a fraction $f\sim 0.1-1$
of the volume;
$f=1$ corresponds to a homogeneous mixture of cool and hot 
phases. The  fraction $f$ should not be confused with
the filling factor of the mixed cool dense gas,
which is of the order of $10^{-5}$.
Observationally, the lumpiness of mixing zones
could be manifested as a veiling effect, i.e., a comparable moderate
depth of absorption lines with  different optical depth
(viz., H$\alpha$ and H$\beta$ as in our case).
To facilitate the treatment of the radiation transfer in the lumpy medium
we use an occultation optical depth,
$\tau_{\rm oc}$, i.e., the average number of lumps along the radius.
This value is defined by $f$ and a ratio
of the total thickness of the mixing layer (forward post-shock
layer $\Delta R$) to the mean intercepted length of the lump $d$
(for a sphere, $d={4\over 3}\times$radius):
$\tau_{\rm oc}=f(\Delta R/d)$. Generally, $\tau_{\rm oc}$ could be
different for
each of the three components. However, we assume the same value of $\tau_{\rm oc}$
for all three components. The effective optical depth in the
line ($\tau$) then can be expressed
through the average optical depth of homogeneously distributed CDS material
$\tau_{\rm av}$ and $\tau_{\rm oc}$:
$\tau=\tau_{\rm oc}[1-\exp(\tau_{\rm av}/\tau_{\rm oc})]$ \citep{CC06}.
We thus consider a three component model of the cool dense gas 
which includes: (i) a narrow CDS component 
with turbulent velocity $u_{\rm t}\approx 20$ km s$^{-1}$,
(ii) a broad  CDS component with $u_{\rm t}\approx 400-500$ km s$^{-1}$, 
both powered by X-rays, and (iii) a broad CDS component powered 
by the thermal conduction. 

\begin{figure}
\plotone{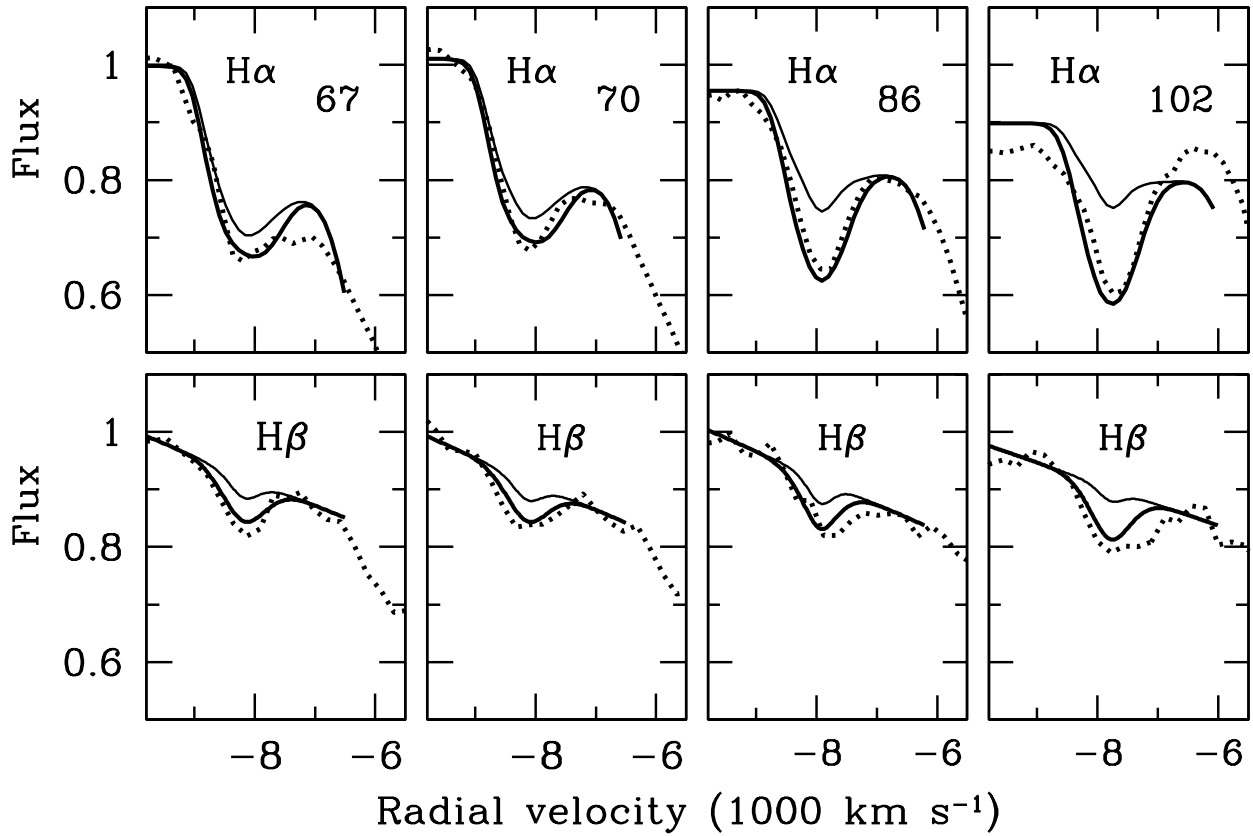}
  \caption{Evolution of the HV absorption in the
H$\alpha$ and H$\beta$ lines of SN~2004dj in a model including
absorption by CDS gas.
The {\em dotted} lines are the observed spectra \citep{Vin06},
the {\em thin} line is the two-component CDS model composed of narrow   and 
broad CDS components; the {\em thick} line is the
three-component CDS model (Table 2) with the  third component being
a broad CDS component powered by thermal conduction.
}
  \label{f-HVAev}
  \end{figure}

The CDS notches in H$\alpha$ and H$\beta$ in SN~2004dj
are calculated on the basis of the previous SN~2004dj model 
assuming
$u_{\rm t}=20$ km s$^{-1}$ for narrow and $u_{\rm t}=400$ km s$^{-1}$
for broad CDS components.  The adopted fraction of the broad CDS component is 
$\beta=0.5$ at the early epoch ($t=67$ d), with a subsequent increase
proportional to the age, while the fraction of the third 
component is $\mu=0.1$ (Table 2). 
The CDS absorption in H$\alpha$ is superimposed on the model 
line profile with the ejecta HV absorption.
For the H$\beta$ line, the CDS HV absorption is superimposed on the 
background spectrum taken as a linear interpolation between 
the blue and red wings of HV absorption in the observed profile.
The occultation optical depth $\tau_{\rm oc}\sim 0.15-0.4$ (Table 2)
is chosen to satisfy the ratio H$\alpha$/H$\beta$, while
the area ratio $C_{\rm s}$ is determined from
the strength of the absorption. Fig. (\ref{f-HVAev}) 
shows that a three-component model with
the area ratio $C_{\rm s}\sim 8-30$ (Table 2) describes
the HV absorption of both H$\alpha$ and H$\beta$ lines satisfactorily. 
On day 102 the ejecta HV absorption is somewhat 
stronger, which is responsible  
for the red-blue asymmetry in the model profile.
However, this does not affect the intensity of the CDS absorption feature.
The comparison to the two-component model and the behavior of $C_{\rm s}$
show that the role of 
thermal conduction increases with time, which is the expected result.
Interestingly, the result depends rather weakly  on $\mu$. This is 
related to the canceling of effects of the variation of $\mu$ in 
the thermal excitation of hydrogen 
and the column density of the third component.
We performed  modeling of the 
CDS HV absorption adopting the lower boundary velocity, 
$v_{\rm c}=11,000$ km s$^{-1}$, and found that in this case 
deceleration of the CDS is too slow, which leads to a 
larger velocity of the CDS HV absorption on day 102 than is observed. 
The value $v_{\rm c}=13,000$ km s$^{-1}$ is about the least that 
provides the required deceleration rate of the CDS. 

Although the parameter values used for the third component are reasonable,
the modeling cannot be considered as proof that thermal
conduction actually dominates the excitation of
the CDS material at a late epoch -- the model contains many assumptions.
Yet the need for an additional source of excitation of hydrogen
responsible for the HV absorption is an established result. Thermal conduction is a natural
mechanism strongly suggested by the inevitable RT instability and
mixing of the CDS matter with hot gas from the forward shock.
On the other hand, the
additional excitation might 
at least partially be related to enhanced soft X-ray radiation from hot dense inhomogeneities produced by the RT mixing 
and heated by thermal conduction up to $\sim10^{(6-7)}$~K.


\subsection{CDS absorption as a wind probe}\label{sec-probe}

The identification of the features produced by  
H$\alpha$ and H$\beta$ absorption in the CDS provides
an efficient tool for the direct determination of the 
velocity of the SN/wind interface in SNe~IIP.
With this value in hand we are able to obtain
the wind parameter $w$ using the thin shell
model of the deceleration. Of course, the method
assumes that the ejecta parameters $M$, $E$, and $k$ are known.

\begin{figure}
\plotone{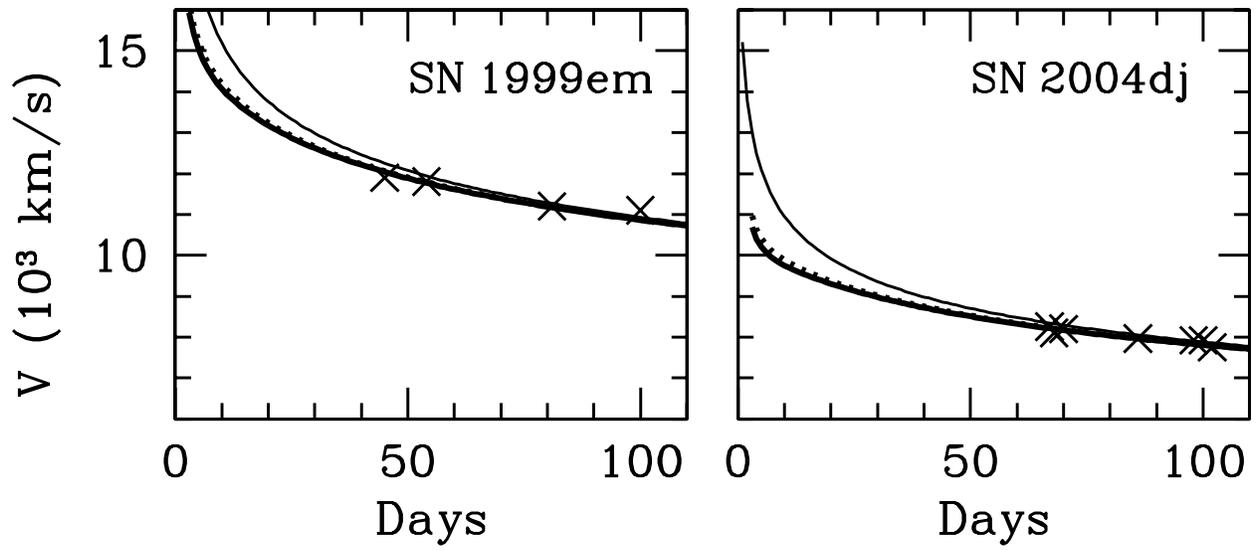}
  \caption{Velocity of the HV absorption in H$\alpha$ of SN~1999em and SN~2004dj.
The {\em crosses} are the velocities of the HV absorption from
observed spectra; the {\em thick solid} and {\em dotted}
lines are the velocities of the HV absorption with and without limb darkening 
in the self-similar model;
the {\em thin solid} line is the velocity of the CDS in the self-similar model.
}
  \label{f-cds}
  \end{figure}

To facilitate the application of this method we use
the self-similar solution. This description
is valid at the late photospheric epoch, when the
initial conditions related to the structure of
the outermost ejecta layers are already `forgotten.'
The self-similar evolution of the CDS radius suggests
$R=At^{\omega}$, with $\omega=(k-3)/(k-2)$ \citep{Che82a,Che82b,Nad85}. 
The factor $A$ for the adopted density distribution 
(eq. [\ref{eq-den}]) is
\begin{equation}
A=\left[\frac{2Mv_0^{(k-3)}}{(k-4)(k-3)wC_m }\right]^{1/(k-2)}\,,
\end{equation}
were $C_m$ is defined by equation (\ref{eq-cmce}) and $v_0$ by equation (\ref{eq-v0}).
The parameters for SN~1999em are as before:
$M=19~M_{\odot}$, $E=1.3\times10^{51}$ erg, and $k=9$.
The radial velocity of HV absorption is calculated both for
the Lambert photosphere (eq. [\ref{eq-hva}]) and with limb darkening
described by the first Chandrasekhar approximation. The
evolution of the photospheric velocity is set according to \cite{Leo02}.
The observational estimates of the radial velocity of HV absorption of H$\alpha$ are
obtained using the spectra reported by \cite{Leo02}.
The best fit for the wind density parameter is found to be
$w=0.99\pm0.09$ (Fig. \ref{f-cds}).
This estimate is in a good agreement with the value derived from
the ejecta HV absorption of H$\alpha$ ($w=1.00\pm0.05$) and 
the He\,I 10830 \AA\ line ($w=1.15\pm0.07$).
Uncertainties in the limb darkening law and in the
photospheric velocity do not affect the derived value of the wind density.

In the case of SN~2004dj we use the same model as before:
 $M=19~M_{\odot}$, $E=0.67\times10^{51}$ erg,
and $k=9$. The radial velocities of
HV absorption are measured from spectra obtained by \cite{Vin06}.
The best fit wind parameter is $w=1.35\pm0.05$ (Fig. \ref{f-cds}).
This agrees with the value
$w=1.27\pm 0.1$ obtained for SN~2004dj from the ejecta HV absorption of H$\alpha$.

The coincidence of $w$ values found by diagnostics based on the 
HV absorption in ejecta and in the CDS is remarkable because the 
first probe is based on the model of X-ray emission by the 
reverse shock and the model of 
the ionization and excitation of ejecta, while the second is based 
only on the self-similar dynamics of the thin shell.

\section{DISCUSSION AND CONCLUSION}

Our goal was to study signs of CS interaction in the
H$\alpha$ and He\,I 10830 \AA\ absorption lines of SN~IIP during the
photospheric stage. We developed a model for the ionization and excitation of
 H and He in the unshocked ejecta taking into account
time-dependent effects and the irradiation of ejecta by X-rays.
The principal result 
is a prediction that for a typical RSG wind ($w\sim1$),
the effects of CS interaction
in a normal SN~IIP are significant and can be detectable
at the photospheric stage in both lines.
These effects consist of the emergence of: (i) HV absorption in the blue wing
of the undisturbed H$\alpha$ line (absorption `shoulder')
at $t\sim40-80$ d, and (ii) HV absorption in He\,I 10830 \AA\ at an
age $t\sim20-60$ days at a radial velocity
of $\sim 10^4$ km s$^{-1}$.
We identify the HV absorption in SN~1999em 
in both H$\alpha$ and He\,I 10830 \AA\ and derive a wind density parameter
$w\approx1$.
HV absorption in H$\alpha$ is identified in SN~2004dj as well and the value of the wind density parameter
is again found to be $w\approx1$.

The second important finding is that, in addition to the ejecta,  
absorption in the CDS formed at the SN/wind interface
plays an important role in the HV feature, especially  
at the late photospheric phase.
This component of the HV absorption is manifested 
as a notch in the HV absorption of H$\alpha$ in
SN~1999em and SN~2004dj after about day 60. A similar 
notch is seen in the H$\beta$ line of these SNe.
We developed a model that produces the notch in both 
H$\alpha$ and H$\beta$ at the different epochs of SN~2004dj 
which suggests a growing role for thermal conduction in the 
heating of the mixed CDS gas.
The wind densities estimated from the observed 
velocities of HV absorption produced by the CDS 
in SN~1999em and SN~2004dj are consistent with the densities 
derived from the HV absorption produced by unshocked ejecta.

The identification of  HV absorption in H$\alpha$ and H$\beta$ in 
SN~1999em has been discussed by \cite{Leo02}. 
Here, we identified HV absorption in H$\alpha$ and H$\beta$ in 
SN~2004dj at a velocity $\sim 3000$ km s$^{-1}$ lower than in SN~1999em.
The difference provides an additional argument against a possible
metal line origin for these features.

\cite{Leo02} 
claim  that a similar component is also seen in the Na\,I doublet.
Indeed, there is a weak dip with a depth of $\sim0.02$ at the right velocity,
$\sim -11,500$ km s$^{-1}$, if it is associated with the
Na\,I 5890 \AA\ doublet.
We calculated the optical depth of the Na\,I 5890 \AA\ doublet for SN~1999em
on day 60 in the model of absorption by the CDS and
found that the narrow CDS component in this line has an optical depth
of $\sim0.1$, which means that after broadening by about 10 times
the depth of absorption should become equal to $\sim0.01$, comparable
to the observed notch.
However, we did not find a similar notch in the Na\,I line of 
SN~2004dj, or in any other SN~IIP. At present, 
the reality of the HV absorption in the Na\,I 5890 \AA\ doublet remains
uncertain.

We examined other published spectra of SNe~IIP at the photospheric epoch and
found evidence for  HV absorption in the H$\alpha$ blue wing
of SN~1985P  \citep{CC87}, in both
 H$\alpha$ and He\,I 10830 \AA\ of SN~1995V \citep{Fa98}, in H$\alpha$ of
SN~2003gd, and SN~2006ov \citep{LW07}. 
Available spectra of SN~1999gi do not cover the range
$50-70$ days when the most pronounced CS interaction effects 
should be observed.
In noisy spectra on day 39 and 89 reported by \cite{LFW02},
we do not find unequivocal evidence for HV absorption in H$\alpha$.
We checked spectra of SN~2004et \citep{Sah06} and did not find convincing 
evidence for   HV absorption in H$\alpha$ or  
H$\beta$.   Early stage ($20-40$ d) spectra 
of SN~2004et show a shallow absorption  in the H$\alpha$ blue wing
rapidly evolving towards the red. However the Si\,II 6347, 6371 \AA\ doublet may  
be responsible for this absorption. At a later epoch, a weak notch at 6282 \AA\ 
is present but unfortunately it coincides with the 
telluric absorption 6281.7 \AA\ line
and the identification with HV absorption is doubtful. 
\cite{Sah06} consider the notch in H$\alpha$ of SN~2004et 
to be real and remark that it might be related to CS interaction. 

\begin{figure}
\plotone{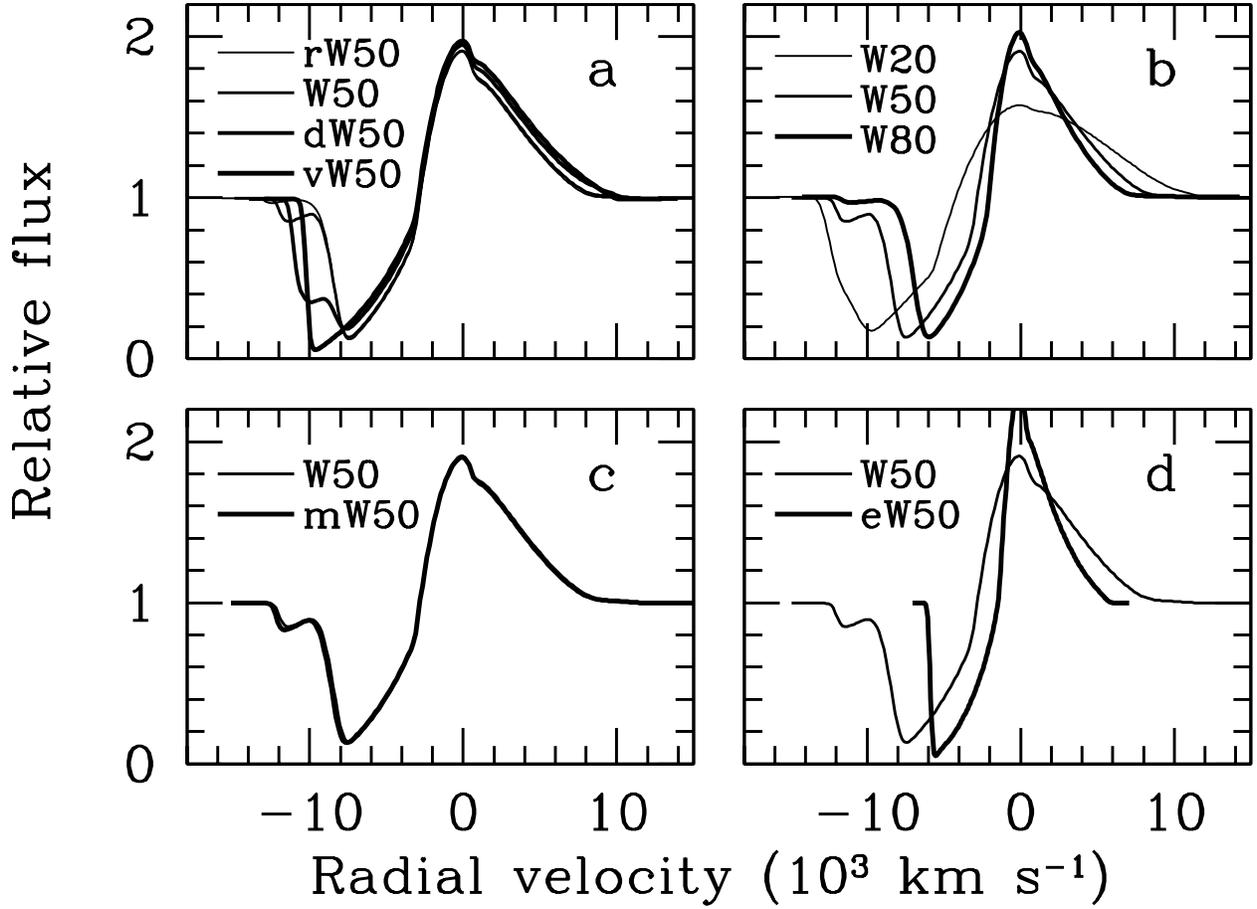}
\caption{The HV absorption in the H$\alpha$ profile for different models
(cf. Table 1). Panel {\em a} shows the effect of wind density on day 50;
panel {\em b} the evolution of the HV absorption in the standard model W on
days 20, 50, and 80; panel {\em c} shows the negligible 
effect of  low ejecta density for the 
low mass case mW50; panel {\em d} shows the interaction effect in
low energy SNe~IIP (cf. Table 1).
}
  \label{f-modha}
  \end{figure}

If the HV absorption in SN~2004et is absent, the question arises 
why the dense wind suggested by a relatively high radio luminosity 
of this SN compared to SN~2004dj and SN~1999em \citep{CFN06} is not 
revealed optically. The answer may be that the wind is so dense that 
the HV absorption is strong and gets merged with the main absorption as in 
the model case of $w=4$ (Fig. \ref{f-modha}). This conjecture is 
consistent with the strong H$\beta$ absorption observed 
in SN~2004et \citep{Sah06}.
A direct test of the proposed explanation might be the detection of 
strong He\,I 10830 \AA\ absorption in SN~2004et. 
Unfortunately, to our knowledge, observations of the He\,I 10830 \AA\ region 
in SN~2004et are lacking.
Interestingly, SN~2006my \citep{LW07}, which does not show HV absorption, 
also has  strong H$\beta$ like SN~2004et. If the same mechanism 
(dense wind, $w\gg1$) explains the absence of the HV absorption in this case, 
then SN~2006my should demonstrate  high radio and X-ray luminosities for a SN~IIP.

We did not find signatures of  HV absorption in H$\alpha$ observations
of the low-luminosity SN~1999br in spectra taken during the first
42 days after the explosion \citep{Pas04}. In the spectrum on day 42,
the H$\beta$ line is rather weak and the explanation that has been invoked for
SN~2004et (strong CS interaction effect) is not applicable  
in this case. 
The likely reason for the absence of HV absorption in SN~1999br is
a low wind density, $w<1$. 
This conjecture has an interesting implication.
Since the mass loss rate increases with the stellar mass according to 
the scaling law of \cite{NdJ90}, we expect that 
the main-sequence mass of pre-SN~1999br should  be lower than that 
of SN~1999em.

An interesting result, although model dependent, is that at the late
photospheric stage of SN~2004dj ($t\sim100$ d) thermal
conduction may be the dominant heating mechanism of the
mixed CDS gas in the forward shock. In this regard,
we recall the H$\alpha$ problem in SN~1979C: at an age
of $\sim1$ yr the H$\alpha$ luminosity was $\sim0.5$ of the total
reverse shock luminosity \citep{CF85}, which is beyond reasonable
values of the H$\alpha$ emission efficiency.
A natural solution to the problem of the high
H$\alpha$ luminosity in SN~1979C could be 
thermal conduction in the mixing layer of the forward shock \citep{Chu97}.
Another possibility, however, is that the mixing process could bring about
 enhanced X-ray emission from the hot dense mixed component, which
may be another source of additional excitation of hydrogen in the CDS.
This soft X-ray component might explain 
the softening of the X-ray emission observed in SN~1999em \citep{Poo02}.

Using {\em Chandra} X-ray data,  \cite{Poo02} conclude that the wind
density parameter for SN~1999em is in the range of $w\sim1-2$
(or $1.5-3$ for the distance 11.7 Mpc).
Our present value ($w\approx1$) is within a factor of 1.5
of the lower limit.
This should be considered as  good agreement for independent
estimates. For SN~2004dj the X-ray data are missing.
The interpretation of radio data for SN~1999em and SN~2004dj
in the model of  wind free-free absorption indicates that the wind density
in SN~2004dj is twice as low as in SN~1999em \citep{CFN06}.
This is somewhat discrepant with the present results  which
indicate similar wind densities for these supernovae.
The
wind density of SN~1999em may be 
overestimated compared to SN~2004dj from radio data
because of the uncertain maximum in the 8.47 GHz light curve or a
possible contribution of synchrotron self-absorption in SN~1999em \citep{CFN06}.

The wind density around SN~1999em and SN~2004dj
suggests that the amount of the pre-SN material lost at the RSG phase
($\sim10^6$ yr) is moderate in these cases.
Assuming the wind velocity of a typical RSG, $u=10$ km s$^{-1}$, we find with
$w=1$ that at the
RSG stage the mass lost by the RSG wind is
$\sim 1~M_{\odot}$. With the ejecta mass of $19~M_{\odot}$
\citep{U07},  $1.4~M_{\odot}$ enclosed in the neutron star, and $<1~M_{\odot}$
lost by the main-sequence wind, the initial mass
of SN~1999em progenitor turns out to be about $22-23~M_{\odot}$.

The mass loss rate by massive stars is often estimated using the formula of
\cite{NdJ90} in which $\dot{M}$ is defined 
through the stellar luminosity ($L$), mass ($M$), and radius ($R$).
For the RSG stage, this formula can be modified using the relation between
the mass and luminosity $L/L_{\odot}=74(M/M_{\odot})^{2.523}$ obtained
from evolution calculations of \cite{MM03} in the range
$15-25~M_{\odot}$ for moderate/no rotation. With this relation,
the mass loss rate by \cite{NdJ90} becomes
$\dot{M}=2\times10^{-12}(M/M_{\odot})^{3.29}(R/R_{\odot})^{0.81}~M_{\odot}$ yr$^{-1}$.
Since the luminosity at the RSG stage is determined by the mass of the
He core which in turn is determined by the initial mass, we
should use the initial mass to estimate $\dot{M}$. For SN~1999em
with $M=22~M_{\odot}$ and $R=500~R_{\odot}$ \citep{U07}
this formula predicts a mass loss rate
$\dot{M}\approx8\times10^{-6}~M_{\odot}$ yr$^{-1}$, i.e., eight times
larger than our value. A similar result is expected for SN~2004dj.
The disparity between our estimate of the pre-SN mass loss
rate and the popular scaling law requires an explanation.
The disparity cannot be attributed to a lower metallicity, because
the metallicities of NGC 1637 (host galaxy for SN~1999em) and NGC 2403
(host galaxy for SN~2004dj) are approximately solar \citep{PVC04}.
The difference can be reduced by a factor of 1.5 if we adopt a higher
velocity for the pre-SN wind, $u=15$ km s$^{-1}$ instead of 10 km s$^{-1}$.
The choice of $15$ km s$^{-1}$ is supported by the
wind velocity of the massive RSG Betelgeuse, $u=14.3$ km s$^{-1}$ \citep{Mau90}.
This, however, does not completely resolve the problem. 
We admit the possibility that the 
hydrodynamical model of the light curve \citep{U07} 
might significantly overestimate
the mass of ejecta for some unknown reason. In that case, to 
obtain the wind $w\approx 1$ (for $u=15$ km s$^{-1}$) the initial mass of the
presupernova must be $M=13.5~M_{\odot}$ with an ejecta mass of 
$M\approx11~M_{\odot}$ instead of $M\approx19~M_{\odot}$. 
With the SN energy reduced accordingly, the depth and position of the 
HV absorption remain the same (Fig. \ref{f-modha}{\em c}).

To conclude, we have found three new probes of the wind density
in SNe~IIP that rely on spectroscopic observations during the photospheric
epoch 20-100 d of (1) the ejecta HV absorption
in H$\alpha$; (2) the ejecta HV absorption in He\,I 10830 \AA; and 
(3) the HV absorption (notch) produced by the CDS. 
The third method is  more practical than the other two because
it does not require  modeling of the absorption lines; it
uses velocity measurements and the self-similar model for the
velocity fit. Of course, this method, as well as the other two,
assumes that we know the ejecta parameters ($M$, $E$, $k$).
However, the application of all three methods may be used to constrain
ejecta parameters as well. The first two methods are sensitive
to 
winds with a density parameter $w>0.5$. However in H$\alpha$ 
CS interaction effects saturate for large density $w\ga4$.
Therefore, only He\,I 10830 \AA\ remains a useful diagnostic 
tool for a high density wind with $w>2$. For the third method 
the range of accessible wind density is less certain, because
a CDS formed by the initial boundary shell
could exist in principle even in the case of a low density wind, although the
hydrogen excitation and shell density  depend on the wind density.
Further 
 understanding of the relation between the strength of deep HV absorption and 
the wind density
requires 3D hydrodynamical modeling of the shock breakout phase in the 
presence of a wind and new spectroscopic observations  in both H$\alpha$ and 
He\,I 10830 \AA\ lines.

\acknowledgements
We thank Daniela Korcakova for sending us spectra of SN~2004dj,
Jozsef Vinko for informing us about SN~2004dj spectra in the
archive SUSPECT,  D. K. Sahu for sending us spectra of SN~2004et,
and the referee for helpful comments on the manuscript. 
This research was supported in part by NSF grant AST-0307366.


{}


\clearpage
\begin{table}
  \caption{Parameters for the models of HV absorption}
  \bigskip
  \begin{tabular}{lccc}
  \hline

Model & $M$ & $E$     & $w$ \\

      &  $M_{\odot}$    &  $10^{51}$ erg  &    \\

\hline

rW    & 19  & 1.3   &   0.5      \\
W     & 19  & 1.3   &   1       \\
dW    & 19  & 1.3   &   2        \\
vW    & 19  & 1.3   &   4        \\
mW    & 10  & 0.9   &   1        \\
eW    & 19  & 0.3   &   1        \\

\hline
\end{tabular}
\label{t-par1}
\end{table}


\begin{table}
  \caption{Model parameters for the notch in SN~2004dj}
  \bigskip
  \begin{tabular}{lcccc}
  \hline

Day & $\beta$   & $\mu$ &  $C_{\rm s}$ &  $\tau_{\rm oc}$ \\

\hline

67  &    0.5    &  0.1     &  8       &  0.15  \\
70  &    0.5    &  0.1     &  8       &  0.15 \\
86  &    0.6    &  0.1     &  13      &  0.35 \\
102  &   0.7     & 0.1     &  30      &  0.4 \\

\hline
\end{tabular}
\label{t-par2}
\end{table}

\clearpage

\clearpage

\clearpage

\clearpage

\clearpage

\clearpage

\clearpage

\clearpage

\clearpage

\clearpage

\clearpage

\clearpage

\end{document}